\documentclass[final,conference]{IEEEtran}

\usepackage[utf8]{inputenc}
\usepackage[T1]{fontenc}
\usepackage{amsmath}
\usepackage[english]{babel}
\usepackage[table]{xcolor}
\usepackage{graphicx}
\usepackage{textcomp}
\usepackage{listings}
\usepackage{cite}
\usepackage{enumerate}
\usepackage{caption}
\usepackage{subcaption}
\usepackage[]{algorithm2e}

\usepackage[normalem]{ulem}

\usepackage[colorinlistoftodos,prependcaption,textsize=tiny]{todonotes}

\usepackage{color}

\definecolor{dkgreen}{rgb}{0,0.6,0}
\definecolor{gray}{rgb}{0.5,0.5,0.5}
\definecolor{mauve}{rgb}{0.58,0,0.82}

\newcommand{\ignore}[1]{}

\begin{document}

\captionsetup[figure]{labelfont={bf,it},textfont=it}

\title{DXRAM's Fault-Tolerance Mechanisms Meet High Speed I/O Devices}
\author{
	\IEEEauthorblockN{Kevin~Beineke, Stefan~Nothaas and Michael~Schöttner}\\
	\IEEEauthorblockA{Institut für Informatik, Heinrich-Heine-Universität Düsseldorf,\\Universitätsstr. 1, 40225 Düsseldorf, Germany\\E-Mail: Kevin.Beineke@uni-duesseldorf.de}
}
\maketitle

\begin{abstract}
In-memory key-value stores provide consistent low-latency access to all objects which is important for interactive large-scale applications like social media networks or online graph analytics and also opens up new application areas. But, when storing the data in RAM on thousands of servers one has to consider server failures. Only a few in-memory key-value stores provide automatic online recovery of failed servers. The most prominent example of these systems is RAMCloud. Another system with sophisticated fault-tolerance mechanisms is DXRAM which is optimized for small data objects. In this report, we detail the remote replication process which is based on logs, investigate selection strategies for the reorganization of these logs and evaluate the reorganization performance for sequential, random, zipf and hot-and-cold distributions in DXRAM. This is also the first time DXRAM's backup system is evaluated with high speed I/O devices, specifically with 56 GBit/s InfiniBand interconnect and PCI-e SSDs. Furthermore, we discuss the copy-set replica distribution to reduce the probability for data loss and the adaptations to the original approach for DXRAM.
\end{abstract}


\section{Introduction}
\label{introduction}
In \cite{dxram4} and \cite{dxram5}, we described the distributed logging and highly parallelized recovery approaches of DXRAM. While we demonstrated that DXRAM outperforms state-of-the-art systems like RAMCloud, Aerospike or Redis on typical cluster hardware, we could not explore the limits of DXRAM's logging approach because of hardware limitations. In this report, we evaluate the backup system of DXRAM with fast hardware and present three different optimizations: (1) an improved pipeline from network to disk on backup servers, (2) a new segment selection strategy for the reorganization of logs and (3) an adapted copy-set approach to decrease the probability for data loss.

The evaluation shows that DXRAM is able to log more than 4,000,000 64-byte chunks per second received over an InfiniBand network. Larger chunks, e.g., 512-byte chunks, can be logged at nearly 1 GB/s, saturating the PCI-e SSD. The reorganization is able to keep the utilization under 80\% most times for all update distributions (sequential, random, zipf and hotNcold) while maintaining a high logging throughput. Furthermore, we show that the two-level logging concept improves the performance up to more than nine times. 

The structure of this report is as follows. Section \ref{dxram} outlines the basic architecture of DXRAM. In Section \ref{related_work_logging}, we depict the related work on logging. In Section \ref{logging_overview}, we give a top-down overview of DXRAM's logging followed by a detailed description of the logging pipeline. Section \ref{related_work_segments} discussed the related work regarding the log reorganization and segment selection. DXRAM's reorganization approach is presented in Section \ref{segments}. The related work for backup distribution is outlined in Section \ref{related_work_copysets} which is followed by the modified copyset approach of DXRAM in Section \ref{copysets}. In Section \ref{evaluation}, we evaluate the proposed concepts of Sections \ref{logging_details} and \ref{segments}. Section \ref{conclusions} concludes this report.

\section{DXRAM}
\label{dxram}
DXRAM is an open source distributed in-memory system with a layered architecture, written in Java \cite{dxramGit}. It is extensible with additional services and data models beyond the key-value foundation of the DXRAM core. In DXRAM, an in-memory data object is called a \emph{chunk} whereas an object stored in a log on disk is referred to as \emph{log entry}. The term \emph{disk} is used for Solid-State Drives (SSD) and Hard Disk Drives (HDD), interchangeably.

\subsection{Global Meta-Data Management}
\label{meta-data}
In DXRAM, every server is either a peer or a superpeer. Peers store chunks, may run computations and exchange data directly with other peers, and also serve client requests when DXRAM is used as a back-end storage. Peers can be storage \textit{servers} (with in-memory chunks), \textit{backup servers} (with logged chunks on disk) or both. Superpeers store global meta-data like the locations of chunks, implement a monitoring facility, detect failures and coordinate the recovery of failed peers, and also provide a naming service. The superpeers are arranged in a zero-hop overlay which is based on Chord \cite{chord1} adapted to the conditions in a data center. Moreover, every peer is assigned to one superpeer which is responsible for meta-data management and recovery coordination of its associated peers. During server startup, every server receives a unique node ID.

Every superpeer replicates its data on three succeeding superpeers in the overlay. If a superpeer becomes unavailable, the first successor will automatically take place and stabilize the overlay. In case of a power outage, the meta-data can be reconstructed based on the recovered peers' data. Thus, storing the meta-data on disk on superpeers is not necessary.

Every chunk in DXRAM has a 64-bit globally unique chunk ID (CID). This ID consists of two separate parts: a 16-bit node ID of the chunk's creator and a 48-bit locally unique sequential number. With the creator's node ID being part of a CID, every chunk's initial location is known a-priori. But, the location of a chunk may change over time in case of load balancing decisions or when a server fails permanently. Superpeers use a modified B-tree \cite{indices}, called \textit{lookup tree}, allowing a space efficient and fast server lookup while supporting chunk migrations. Space efficiency is achieved by a per-server sequential ID generation and ID re-usage in case of chunk removals allowing to manage chunk locations using CID ranges with one entry for a set of chunks. In turn, a chunk location lookup will reply with a range of CIDs, not a single location, only. This reduces the number of location lookup requests. For caching of lookup locations on peers, a similar tree is used further reducing network load for lookups.

\subsection{Memory Management}
\label{memory_management}
The sequential order of CIDs (as described in section \ref{meta-data}) allows us to use compact paging-like address translation tables on servers with a constant lookup time complexity. Although, this table structure has similarities with well known operating systems' paging tables we apply it in a different manner. On each DXRAM server, we use the lower part (LID) of the CID as a key to lookup the virtual memory address of the stored chunk data. The LID is split into multiple parts (e.g., four parts of 12 bit each) representing the distinct levels of the paging hierarchy. This allows us to allocate and free page tables on demand reducing the overall memory consumption of the local meta-data management. Complemented with an additional level indexed by node ID storing of migrated chunks is possible as well. DXRAM uses a tailored memory allocator with very low footprint working on a large pre-reserved memory block. For performance and space efficiency reasons, all memory operations are implemented using the Java Unsafe class.

Chunks store binary data and each chunk ID (CID) contains the \textit{creator}. Chunks can be migrated to other servers for load balancing reasons. Migrated chunks are then called \textit{immigrated chunks} on the receiver and \textit{emigrated chunks} on the creator. Finally, there are \textit{recovered chunks} stored on a new \textit{owner} after a server failure.

\section{Related Work on Logging}
\label{related_work_logging}
Numerous distributed in-memory systems have been proposed to provide low-latency data access for online queries and analytics for various graph applications. These systems often need to aggregate many nodes to provide enough RAM capacity for the exploding data volumes which in turn results in a high probability of node failures. The latter includes soft- and hardware failures as well as power outages which need to be addressed by replication mechanisms and logging concepts storing data on secondary storage. Because of space constraints, we can only discuss the most relevant work.

\textbf{RAMCloud} is an in-memory system, sharing several objectives with DXRAM while having a different architecture, providing a table-based in-memory storage to keep all data always in memory. However, the table-based data model of RAMCloud is designed for larger objects and suffers from a comparable large overhead for small data objects \cite{dxram2}. It uses a distributed hash table, maintained by a central coordinator, to map 64-bit global IDs to nodes which can also be cached by clients. DXRAM on the other hand uses a superpeer overlay with a more space-efficient range-based meta-data management.
For persistence and fault tolerance it implements a log-based replication of data on remote nodes' disks \cite{ramcloudLogging}. In contrast to other in-memory systems, RAMCloud organizes in-memory data also as a log which is scattered for replication purposes across many nodes' disks in a master slave coupling \cite{ramcloudScalable}. Scattering the state of one node's log on many backup nodes allows a fast recovery of 32 GB of data and more. 
Obviously, logging throughput depends on the I/O bandwidth of disks as well as on the available network bandwidth and CPU resources for data processing. RAMCloud uses a centralized log-reorganization approach executed on the in-memory log of the server which resends re-organized segments (8 MB size) of the log over the network to backup nodes. As a result, remaining valid objects will be re-replicated over the network after every reorganization iteration to clean-up the persistent logs on remote nodes. This approach relieves remote disks but at the same time burdens the master and the network.
DXRAM uses an orthogonal approach by doing the reorganization of logs on backup nodes avoiding network traffic for reorganization. \ignore{We also argue that we leave it up to backup nodes to decide when they have free resources to run a reorganization and not burden nodes and network in case of high load.} Furthermore, DXRAM does not organize the in-memory storage as a log but uses updates in-place. \ignore{Thus the need to run a compactification is very unlikely and only for changing access patterns. Another difference is the two stage logging approach in DXRAM which is designed for small data objects as well as improving recovery speed by providing logs sorted per master.}
Finally, RAMCloud is written in C++ and provides client bindings for C, C++, Java and Python \cite{ramcloudLogging} whereas DXRAM is written in Java.

\textbf{Aerospike} is a distributed database platform providing consistency, reliability, self-management and high performance clustering \cite{aerospike}. Aerospike uses Paxos consensus for node joining and failing and balances the load with migrations. In comparison, DXRAM also offers a migration mechanism for load balancing. The object lookup is provided by a distributed hash table in Aerospike. Aerospike is optimized for TCP/IP. Additionally, Aerospike enables different storage modes for every namespace. For instance, all data can be stored on SSD with indexes in RAM or all data can be stored in RAM and optionally on SSD with a configurable replication factor. As Aerospike is a commercial product, not many implementation details are published except that it internally writes all data into logs stored in larger bins optimized for flash memory. The basic server code of Aerospike is written in C and available clients include bindings for C, C\#, Java, Go, Python, Perl and many more.

\textbf{Redis} is another distributed in-memory system which can be used as an in-memory database or as a cache \cite{redis}. Redis provides a master-slave asynchronous replication and different on-disk persistence modes. To replicate in-memory objects, exact copies of masters, called slaves, are filled with all objects asynchronously. To overcome power outages and node failures, snapshotting and append-only logging with periodical rewriting can be used. However, to replicate on disk the node must also be replicated in RAM which increases the total amount of RAM needed drastically. This is an expensive approach and very different from the one of DXRAM where remote replicas are stored on SSD only. Obviously, Redis has no problems with I/O bandwidth as it stores all data in RAM on slaves and can postpone flushing on disk as needed. Furthermore, reorganization is also quite radical compared to DXRAM as Redis just reads in a full log to compress it which is of course fast but introduces again a lot of RAM overhead. Redis is written in C and offers clients for many programming languages like C, C++, C\#, Java and Go.

\textbf{Log-structured File System}s are an important inspiration for the log-based replication of RAMCloud and DXRAM. A log is the preferred data structure for replication on disk as a log has a superior write throughput due to appending objects, only. But, a log requires a periodical reorganization to discard outdated or deleted objects in order to free space for further write accesses. In \cite{LFS} Rosenblum and Ousterhout describe a file system which is based on a log. Furthermore, a cleaning policy is introduced which divides the log into segments and selects the segment with best cost-benefit ratio for reorganization. DXRAM divides a log into segments as well. However, due to memory constrains the cost-benefit formula is limited to the age and utilization of a segment (more in section \ref{related_work_segments}).

\textbf{Journaling} is used in several file systems to reconstruct corruptions after a disk or system failure. A journal is a log that is filled with meta-data (and sometimes data) before committing to main file system. The advantage is an increased performance while writing to the log as appending to a log is faster than updating in-place but requires a second write access. The to be described two-level logging of DXRAM also uses an additional log to efficiently utilize an SSD. In contrast to journaling, we use this log only for small write accesses from many remote nodes to allow bulk writes without impeding persistence.

\section{Logging in DXRAM - An Overview}
\label{logging_overview}
In this section, we describe the basic logging architecture of DXRAM which is subject of \cite{dxram4}. Below, we distinguish two different roles: \emph{Masters} are DXRAM peers, store chunks (see Section \ref{dxram}) and replicate them on \emph{backup servers}. A backup server might also be a master and vice versa.

Replicating multi-billion small data objects in RAM is too expensive and does not allow to mask power outages. Therefore the backup data structures of DXRAM are designed to maximize throughput of SSDs by using logs. 

\subsubsection{Two-Level Logging}
\label{two-level}
We divide every server's data into backup zones of equal size. Each backup zone is stored in one separate log on every assigned backup server (typically three per backup zone). Those logs are called \emph{secondary logs} and are the final destination for every replica and the only data structure used to recover data from. By sorting backups per backup zone, we can speed-up the recovery process by avoiding to analyze a single log with billions of entries mixed from several masters (as required RAMCloud). The two-level log organization also ensures that infrequent written secondary logs do not thwart highly burdened secondary logs by writing small data to disk and thus utilizing the disk inefficiently. At the same time, incoming objects are quickly stored on disk to sustain power outages.

\begin{figure}[!t]
	\centering
	\includegraphics[width=3.5in]{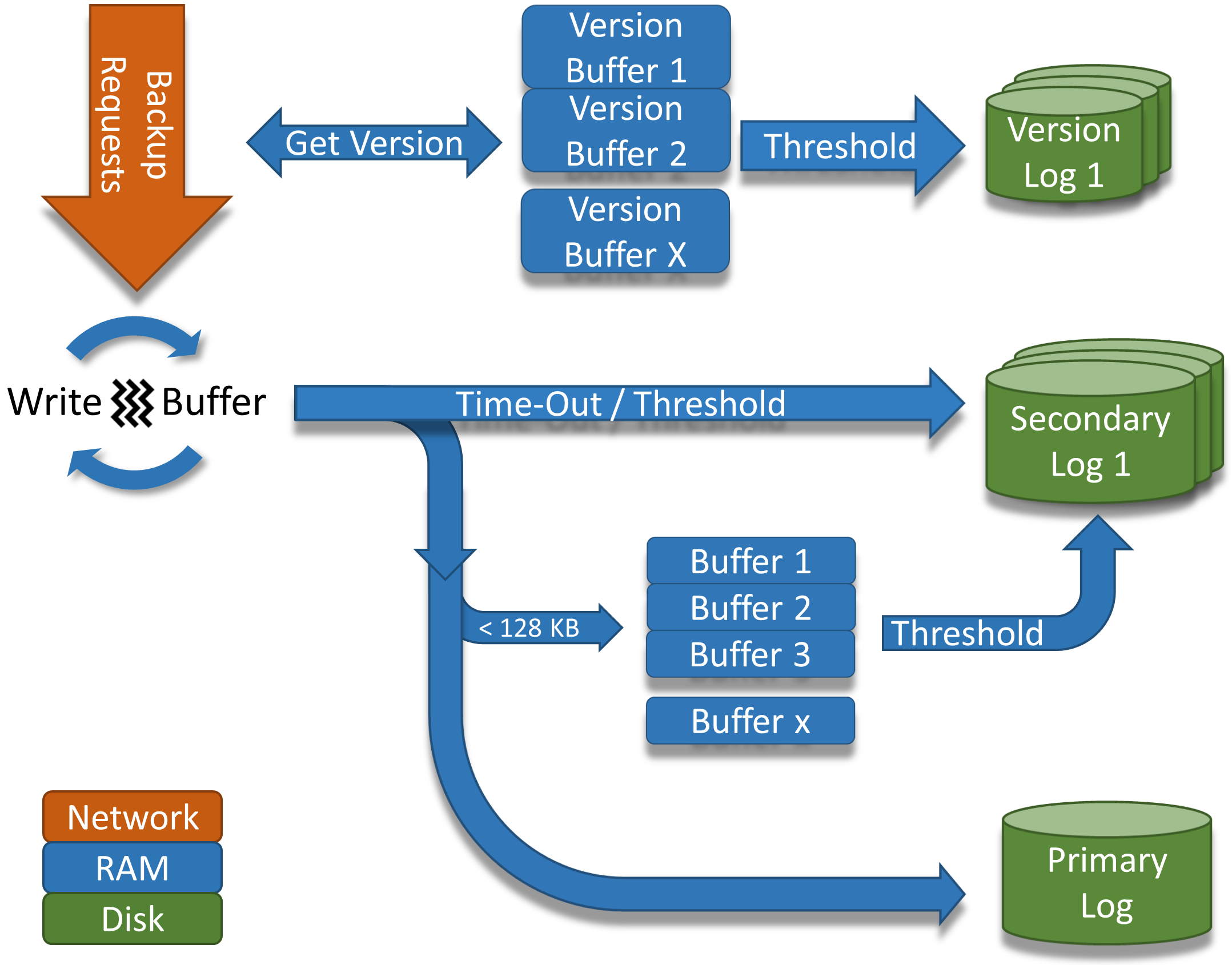}
	\vspace{-15pt}
	\caption{The Logging architecture. Every object is buffered first. Depending on the amount of data per backup zone, the objects are either directly written to their associated secondary log or to both primary log and secondary log once there is enough data. Versions are determined by inquiring the corresponding version buffer, which is flushed to its version log frequently.}
	\label{fig_architecture}
	\vspace{-20pt}
\end{figure}

First, every object received for backup is written to a ring buffer, called \emph{write buffer}, to bundle small request (Figure \ref{fig_architecture}). This buffer is a lock-free ring-buffer which allows concurrently writing into the buffer while it is (partly) flushed to disk. During the flushing process, which is triggered periodically or if a threshold is reached, the content is sorted by backup zones to form larger batches of data in order to allow bulk writes to disk. If one of those batches is larger than a predefined threshold (e.g., 32 flash pages of the disk), it is written directly to the corresponding secondary log.

\begin{figure*}[!t]
	\centering
	\includegraphics[width=6.0in]{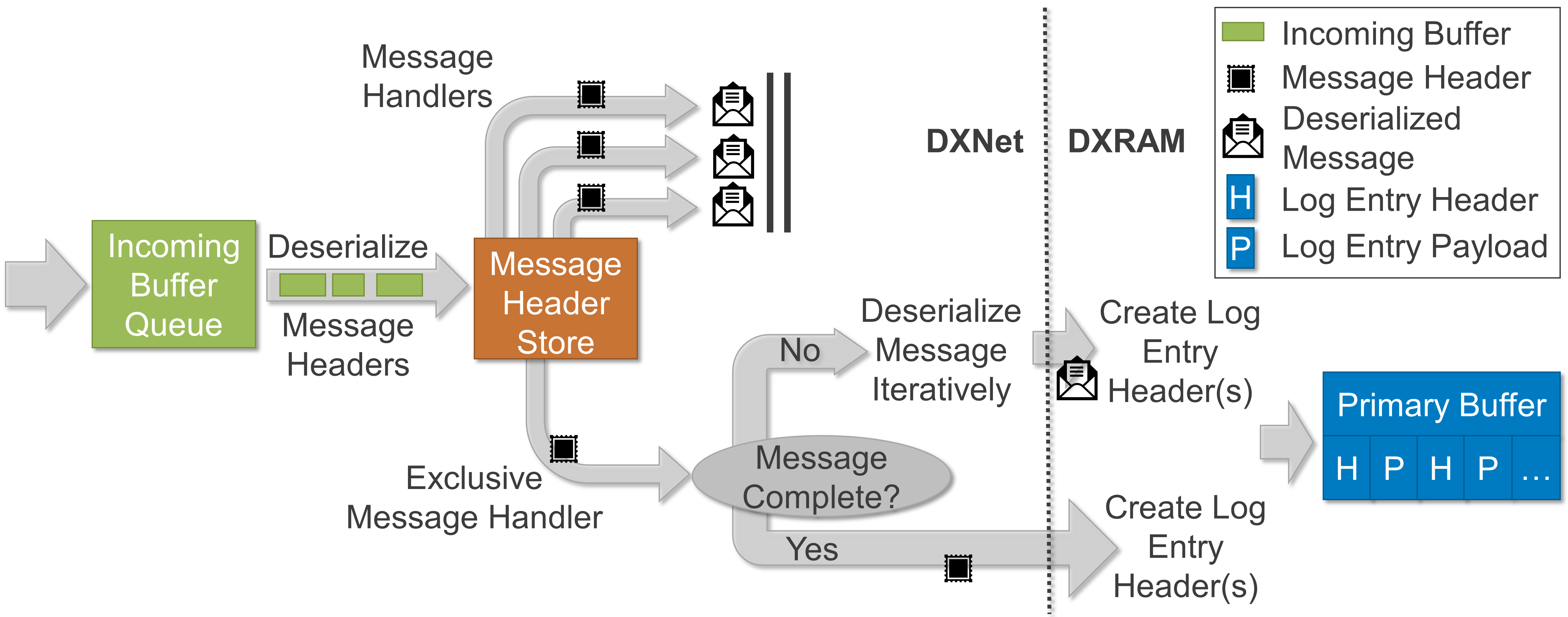}
	\caption{From Network to Write Buffer}
	\label{network_fig}
	\vspace{-20pt}
\end{figure*}

In addition to the secondary logs, there is \textbf{one} \emph{primary log} for temporarily storing smaller batches of \textbf{all} backup zones to guarantee fast persistence without decreasing disk throughput. The smaller batches are also buffered in RAM separately, in so called \emph{secondary log buffers}, for every secondary log and will eventually be written to the corresponding secondary log when aggregated to larger batches. Obviously, with this approach some objects will be written to disk twice but this is outweighed by utilizing the disk more efficiently. Waiting individually for every secondary log until the threshold is reached without writing to the primary log, on the other hand, is no option as the data is prone to get lost in case of a power outage.

\subsubsection{Backup-side Version Control}
Masters do not store version information in RAM. Versions are necessary for identifying outdated data in the logs, so the backup servers employ a version control used for the reorganization and recovery. A naïve solution would be to manage every object's version in RAM on backup servers. Unfortunately, this approach consumes too much memory, e.g., at least 12 bytes (8-byte CID and 4-byte version) for every object stored in a log easily sums up to many GB in RAM which is not affordable. Storing the entire version information on disk is also not practical because of performance reasons as this would require reads for each log write. Caching recent versions in memory could possibly help for some access patterns but for the targeted application domain would either cause many read accesses for cache misses or occupy a lot of memory. Instead, we propose a version manager which runs on every backup server and utilizes one \emph{version buffer} per secondary log. The version buffer holds recent versions for this secondary log in RAM until it is flushed to disk. In contrary to a simple cache solution, DXRAM's version manager avoids loading missing entries from secondary storage by distinguishing time spans, called \emph{epochs}, which serve as an extension of a plain version number. At the beginning of an epoch, the version buffer is empty. If a backup arrives within this epoch, its CID will be added to the corresponding version buffer with version number 0. Another backup for the same object within this epoch will increment the version number to 1, the next to 2 and so on. When the version buffer is flushed to disk, all version information is complemented by the current epoch, together creating a unique version. In the next epoch the version buffer is empty again. An epoch ends when the version buffer reaches a predefined threshold allowing to limit the buffer size, e.g., 1 MB per log. During flushing to disk, a version buffer is compacted resulting in a sequence of (CID, epoch, version)-tuples with no ordering. This sequence is appended to a file on disk, creating a log of unique versions for every single secondary log. We call it a \emph{version log}. Over time, a version log contains several invalid entries which are tuples with outdated versions. To prevent a version log from continuously growing, it is compacted during reorganization.

\section{Logging in DXRAM - From Network to Disk}
\label{logging_details}
In this section, we present all stages involved on a backup server from receiving a chunk over a network connection to writing the chunk to disk. This includes the deserialization of the message object (in Section \ref{deserialization}), the creation of a log entry header to identify a chunk within a log (in Section \ref{log_entry_headers}) and the aggregation of all chunks of all backup zones in the \textbf{write buffer} (in Section \ref{write_buffer}). Furthermore, this section covers the sorting and processing of the write buffer to create large batches which can be written to disk efficiently (in Section \ref{flushing}). After that, we briefly describe all data structures on disk and how they are accessed (in Section \ref{data_structures_on_disk}). Finally, we discuss different disk access methods and describe all three implemented methods thoroughly (in Section \ref{writing_to_logs}).

\subsection{Message Receipt and Deserialization}
\label{deserialization}
For sending replicas, DXRAM uses DXNet, a network messaging framework which utilizes different network technologies, currently supporting Ethernet and InfiniBand. DXNet guarantees packet and message ordering by using a special network handler, which is used for logging. DXRAM defines a fixed replication ordering for every backup zone enabling the application of asynchronous messages for chunk replication. Server failures are handled by re-replicating the chunks to another backup server and adjusting the replication ordering (the failed server is removed and the new backup server is added at the end).

There are two major messages involved in the logging process. One for replicating one or multiple chunks to a specific backup zone (a log message) and one for creating a new backup zone on a backup server which includes creating the secondary and version log and their corresponding buffers. All chunks of one log message belong to the same backup zone (allocation is performed on masters). Therefore, the range ID (identifier for a backup zone which is also called backup range) and the owner is included in the message once, only, followed by the chunk ID, payload length and payload of the first chunk. Typically, messages are created and deserialized entirely by DXNet, i.e., a new message object is created and all chunks are deserialized (in this case into a ByteBuffer) to be processed (logged) by the message handler. For the logging, we optimized this step by deserializing directly into the \textbf{write buffer} (see Section \ref{two-level}) to avoid creating a message object (allocations are rather expensive) and copying from the deserialized ByteBuffer into the write buffer. Whenever a message is contained entirely in the received incoming buffer, the log message is deserialized into the write buffer. If not, i.e., the log message is split into at least two buffers, we delegate the deserialization to DXNet in order to reduce complexity (see Figure \ref{network_fig}). The performance is mostly unaffected by the latter, as log message splitting is rather seldom. The detection is done within DXNet: if the message size is smaller than the number of remaining bytes in the buffer, a special message receiver is called which is registered by the logging component on system initialization. Otherwise, a normal message receiver is called. The difference is that a normal message receiver operates on a deserialized message object and the special message receiver on a message header and not yet deserialized ByteBuffer. We hide the complexity of the deserialization in the special message receiver by using DXNet's Importer which offers deserialization methods for primitives, arrays and objects.

\subsection{Log Entry Headers}
\label{log_entry_headers}

\begin{figure}[!t]
	\centering
	\includegraphics[width=3.5in]{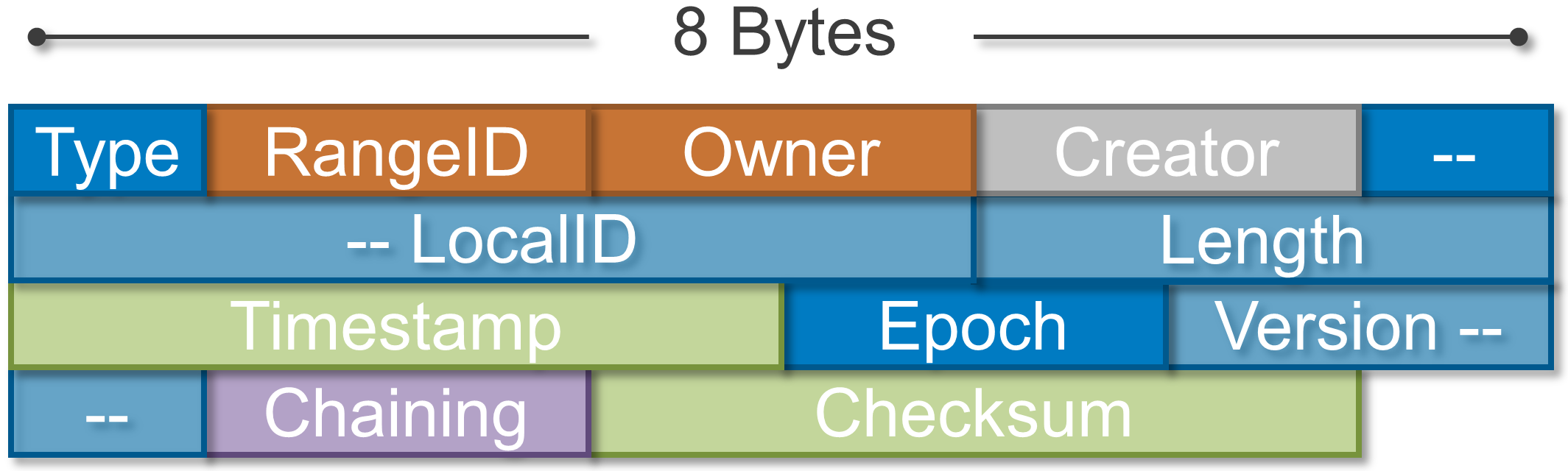}
	\vspace{-10pt}
	\caption{Log Entry Header. Orange: for write buffer and primary log, only. Grey: for migrated or recovered chunks. Green: optional/configurable. Purple: for chunks larger than 4 MB. Dark blue: mandatory, minimal size. Transparent blue: maximum size}
	\label{header_fig}
	\vspace{-20pt}
\end{figure}

We cannot simply copy a chunk (or a batch of chunks) from message buffer to write buffer as every log entry consists of a log entry header followed by the payload. The log entry header has to be created just before writing the entry to the write buffer as it contains a unique version number which has to be determined by inquiring the version buffer of given backup zone. Optionally, the log entry header contains a timestamp and a CRC checksum which have to be recorded/generated as well. Figure \ref{header_fig} shows all fields of a log entry header.

\begin{itemize}
	\item Type: this field specifies the type of the log entry header and stores the sizes of the LocalID, length, version and chaining fields. There are three different types of log entry headers: (1) a \texttt{DefaultSecLogEntryHeader} which is used for chunks stored in a secondary log. (2) A \texttt{MigrationSecLogEntryHeader} for migrated/recovered chunks stored in a secondary log and (3) a \texttt{PrimLogEntryHeader} for log entries stored in write buffer or primary log. In write buffer, every log entry is preceded by a \texttt{PrimLogEntryHeader}. When the log entry is written to primary log, the header remains unchanged. For writing the log entry to secondary log (or secondary log buffer) the \texttt{PrimLogEntryHeader} is converted into a \texttt{Default-} or \texttt{MigrationSecLogEntryHeader} by removing the RangeID and owner fields (both specified by the secondary log the entry is stored in). For converting to a \texttt{DefaultSecLogEntryHeader}, the creator is removed as well (the creator is the same as the creator of the backup zone).
	\item RangeID and Owner: for log entries stored in write buffer and primary log to identify the backup zone the log entry belongs to.
	\item Creator: if the creator differs from the creator of the backup zone (if migrated or recovered), it has to be stored in order to restore the CID during recovery.
	\item LocalID: to identify a chunk within a backup zone. Can be one, two, four or six bytes (defined in type field).
	\item Length: the payload size. Can be zero, one, two or three bytes. Maximum size for a log entry is 4 MB (half the size of a segment which is configurable). Larger chunks are split to several log entries (see Chaining).
	\item Timestamp: the timestamp represents the point in time the log entry was created. More precisely, the seconds elapsed since the log component was created. The timestamp is optional and used for the optimized segment selection of the reorganization (see Section \ref{segments}).
	\item Epoch and Version: together epoch and version describe a unique version number for given CID. The version field can be zero, one, two or four bytes. The most used version $1$ takes no space to store.
	\item Chaining: not available for chunks smaller than 4 MB. Otherwise, the first byte represents the position in the chain and the second byte the length of the chain. Theoretical maximum size for chunks with default configuration: $256 * 4 \text{MB} = 1 \text{GB}$. The segment size can be increased to 16 MB to enable logging of 2 GB chunks which is the maximum size supported by DXRAM.
	\item Checksum: the CRC32 checksum is used to check for errors in the payload of a log entry during the recovery. If available, the checksum is generated using the SSE4.2 instructions of the CPU (see Figure \ref{checksum}).
\end{itemize}

\begin{figure}[!t]
	\begin{lstlisting}[xleftmargin=5.0ex]
uint32_t i = 0;
while (i + 8 <= length) {
	crc = _mm_crc32_u64(crc, *((uint64_t *) &data[i + offset]));
	i += 8;
}

if (i + 4 <= length) {
	crc = _mm_crc32_u32(crc, *((uint32_t *) &data[i + offset]));
	i += 4;
}

if (i + 2 <= length) {
	crc = _mm_crc32_u16(crc, *((uint16_t *) &data[i + offset]));
	i += 2;
}

if (i < length) {
	crc = _mm_crc32_u8(crc, data[i + offset]);
	i++;
}
	\end{lstlisting}
	\caption{Fast Checksum Computation}
	\label{checksum}
	\vspace{-20pt}
\end{figure}

\subsection{Write Buffer}
\label{write_buffer}
The write buffer is a ring-buffer which is accessed by a single producer, the exclusive message handler (see Figure \ref{network_fig}), and a single consumer, the \textit{BufferProcessingThread}. Every chunk to be logged is written to the write buffer first. Beneath the data in the write buffer, we also store backup zone specific information in a hash table. The keys for the hash table are the owner and range ID (combined to an integer) of the backup zone. The value is the length of all current log entries belonging to the backup zone. We use a custom-made hash table based on linear probing as it is faster than Java's hashmap and avoids allocations due to reusing the complete data structure for the next iteration. Access to the hash table is not atomic but must be in sync with the write buffer. Thus, updating the write buffers positions and the hash table is locked by a spin lock, even though the write buffer itself could be accessed lock-free. However, the length information is important to distribute the log entries to segments (see Section \ref{flushing}). We do not wait in case the lock cannot be acquired but try again directly because the critical areas are small as well as the collision probability (\textit{BufferProcessingThread} is in critical area for a short time every 16 MB or 100 ms, see Section \ref{flushing}).

If the write buffer is full, the exclusive message handler has to wait for the \textit{BufferProcessingThread} to flush the write buffer. We use LockSupport.parkNanos(100) which is a good compromise between reducing the CPU load while waiting and being responsive enough. When writing to the write buffer, the overflow needs to be dealt with by continuing at the buffers start position. 

\subsection{Flushing and Sorting}
\label{flushing}
The \textit{BufferProcessingThread} flushes the write buffer periodically (every 100 ms) and based on a threshold (half the size of the write buffer; default: 32 MB). The flushing can be done concurrently to further filling the write buffer once the metadata (front and back pointer and the hash table) has been read and set accordingly. Thus under load, half of the write buffer is written to disk while the other half is filled enabling a constant utilization.

\textbf{Priority Flush:} The flushing can also be triggered by the recovery and reorganization to ensure all relevant data is stored in the corresponding secondary logs. Additionally, whenever a version buffer is full, a priority flush is triggered to flush the version buffer consequently.

The flushing process does not simply write the data as it is to the corresponding secondary logs, but sorts the log entries by backup zone to create the largest possible batches which can be written efficiently to disk. First, we use the information about the total length of a backup zone's data (stored in the hash table) to supply ByteBuffers to store the sorted data. None of the buffers exceeds the size of a segment as we want to write the buffer's content with one write access, if possible. For example, 14 MB of data belonging to one backup zone might be split to two 8 MB buffers. It could also be split to one 8 MB and six 1 MB buffers depending on the available buffers in the buffer pool (described below). All buffers of one backup zone are collected in a Java object which is registered with an identifier for the backup zone (combined range ID and owner) in a hash table similar to the one for recording the lengths. This enables a fast lookup during the sorting process. The ordering within a backup zone is preserved because we iterate the write buffer from back pointer to front pointer and copy the log entries to the corresponding buffers. We do not fill previous buffers to reduce fragmentation, either, because of the ordering (a smaller succeeding log entry might fit in the previous buffer). Again, when copying the log entries the overflow of the write buffer must be considered. Additionally, the log entry headers are truncated when written to the buffers (see Figure \ref{header_fig}).

\subsubsection{Buffer Pool}
To avoid constantly allocating new buffers when sorting the data, we employ a buffer pool which stores buffers in three configurable sizes (e.g., 64 x 0.5 MB, 32 x 1 MB and 8 x 8 MB for 8 MB segments) to support different access patterns. The buffer pool consists of three lock-free multi producer, single consumer ring buffers and buffers are chosen with a best-fit strategy. If all ring buffers run dry, the \textit{BufferProcessingThread} waits for the next buffer being returned. Buffers are returned after they have been written to disk.

The buffers of all backup zones with less than 128 KB (default value) of data are merged and written to primary log. Additionally, the buffers are copied to the corresponding secondary log buffers (with further truncated headers) to enable fast flushing once a buffer is full. For backup zones with more than 128 KB of data, the buffer is directly written to secondary log. It might be necessary to flush the secondary log buffer first (can be merged with buffer if both together are not larger than a segment). The \textit{BufferProcessingThread} does not execute the write accesses to disk, but registers the write access in a lock-free ring-buffer, called \textit{WriterJobQueue}, to allow concurrent sorting/processing of new data while the data is written to disk (very important for synchronous access). The WriterJobQueue is synchronized by using memory fences, only. The jobs are pulled and executed by a dedicated writer thread.

After the flushing, the hash table is cleared and the back pointer of the primary buffer is set to previous front pointer.

\subsection{Data Structures on Disk}
\label{data_structures_on_disk}
Everything DXRAM's backup system writes to disk is arranged in logs. The primary log and secondary logs store replicas, the version logs version information for all logged chunks.

\subsubsection{Primary Log}
\label{primary_log}
The primary log is used to ensure fast persistence for arbitrary access while efficiently utilizing the disk, i.e., if the write buffer stores log entries of many backup zones, all batches may be small and writing them to disk would slow down the disk considerably. Thus, all data is written to primary log with one large access and buffered in corresponding secondary log buffers. If a secondary log buffer is large enough to be written to disk efficiently (default: 128 KB), it will be flushed to secondary log.

The primary log is filled sequentially from beginning (position 0) to the end. It does not get reorganized or compacted in any way, nor is it used to recover from during the online recovery. Its only purpose, is to store small batches persistently to be recovered in case of a power outage, i.e., all servers responsible for at least one backup zone break down and not all secondary log buffers could be flushed prior to the failure.

If the primary log is full, all secondary log buffers are flushed and the position is set to 0. As secondary log buffers are flushed frequently the amount of data to be flushed in this scenario is rather small. Even in worst case scenario, only 128 KB per backup zone needs to be written to disk.

For the two-level logging, we assume the cluster servers do not have non-volatile random access memory (NVRAM) or battery backup. If they utilize NVRAM (with NVRAM we refer to byte-addressable non-volatile memory on main memory layer, not flash memory used in SSDs), logging to disk is still necessary as replicating in NVRAM is too expensive and failed servers may be irreparable. However, the two-level logging is redundant as the write buffer and secondary log buffers can be accessed after rebooting (not implemented). For battery backed-up servers, the primary log is expendable, as well, because all secondary log buffers can be flushed while the server runs on battery (soft shut-down). We provide two options to optimize the logging for NVRAM and battery backup: (1) the threshold to decide whether the data is written to primary log or secondary log can be set to $0$ or (2) the two-level logging can be disabled explicitly. In the first case, all aggregated batches are written directly to secondary log regardless of the size of the bulk. The second option disables the primary log, as well, but still utilizes secondary log buffers if batches are small.

\subsubsection{Secondary Logs}
\label{secondary_logs}
Secondary logs eventually store all log entries on disk and are used for the recovery (online and global shut-down recovery). Secondary logs are subdivided into segments (default size: 8 MB) which is beneficial for the recovery and reorganization to limit the memory consumption by processing the log segment by segment. Furthermore, the segmentation allows reorganizing the parts of the log which are more likely outdated (see Section \ref{segments}). For writing new log entries to the secondary log, the segment boundaries must be respected because log entries must not span over two segments which would add unnecessary complexity to the recovery and reorganization.

As described before, log entries are sorted and copied into buffers with maximum size equal to the segment size. Usually, we write an entire buffer into one segment. When the buffer stores at least 6 MB (75\% of the segment size), we open a new segment and write the buffer to the beginning of the new segment. If not (< 6 MB), we search for a used segment with enough space to write the entire buffer into. When none of the used segments have enough space to hold the buffer, we open a new segment, too. If all segments of a log are already in use, we split the buffer and gradually fill the segments with most free space. This is a compromise between maximum throughput while writing to the log (minimum write accesses, page-aligned access) and maximum efficiency of the reorganization (high utilization of the segments). If the buffer contains more data than there is free space in the secondary log, a high-priority request is registered for reorganizing the complete secondary log and the writer thread waits until the request was handled. If the secondary log's utilization breaches a configurable threshold (e.g., 85\%) after writing to disk, a low-priority reorganization request is registered and the writer thread proceeds.

The presented writing scheme is used whenever the secondary log is not accessed by the reorganization thread. If the reorganization is in progress for the secondary log to write to, an \textbf{active segment} is chosen which can be filled concurrently to other segments being reorganized as it is locked to the reorganization. The active segment is exchanged if it is full (next log entry does not fit), only. Obviously, the currently reorganized segment cannot be used as an active segment. All other segments are free to be chosen. Furthermore, during concurrent access all write accesses to disk have to be serialized to avoid corrupting the file.

Secondary logs are recovered entirely by reading all log entries of the log (segment by segment) and storing the valid (and error-free) log entries in DXRAM's memory management. In order to keep recovery times low and to avoid secondary logs completely filling up, secondary logs have to be reorganized from time to time (see Section \ref{segments}). During the recovery, the reorganization is completely locked to avoid inconsistencies and to allocate all available resources to the recovery.
	
\subsubsection{Version Logs}
\label{version_logs}
Version logs store the version numbers of log entries belonging to the same secondary log. Every version log is supported by a version buffer which holds all current versions of the latest epoch. At the end of every epoch, the version buffer is flushed to version log by appending all version numbers of the version buffer to the end of the version log. An epoch transition is initiated whenever the writer thread writes to a secondary log and the version buffer breached its flushing threshold (e.g., 65\%). If the secondary log is reorganized simultaneously, the version buffer is not flushed but a low-priority reorganization request is registered. Prior to the actual reorganization, the reorganization thread has to read the entire version log and store all version numbers in a hash table which is used to validate the log entries during the reorganization. Then, the hash table is complemented by all version numbers currently stored in the version buffer and the epoch number is incremented. We exploit the situation to compact the version log by writing all version numbers stored in the hash table to the beginning of the version log. This way, we keep the version log small without a dedicated reorganization.

In case of a power failure, a version log might be ahead of its secondary log, i.e., a new version number might have been stored in version log whose corresponding log entry have not been written to secondary log prior to the hard shutdown. Therefore, it is important to not use the version log after a power failure to identify the most recent version of a chunk as this could result in not recovering a chunk at all by rejecting the most recent version in secondary log (which has a lower version number than registered in version log). Instead, we cache all chunks from all segments (in a hash table) and overwrite an entry if the version number is higher. The version log is used to determine deleted chunks, only. For the crash recovery of a single server (or multiple servers with at least one alive replica of every backup zone) and the reorganization, the versions are gathered, first. Then, the version log as well as the primary and secondary logs are flushed, prior to recovering/reorganizing the secondary log. Therefore, the version information cannot be more recent than the data. However, the opposite is possible (data newer than version). We solve this by considering all brand-new log entries (logged during the recovery/reorganization) to be valid, i.e., log entries created in the current epoch are kept (the epoch is not incremented during the recovery/reorganization after reading the versions).
		
\subsection{Access Modes for Writing to Logs}
\label{writing_to_logs}
DXRAM supports three different disk access modes to write to logs (primary, secondary and version logs): (1) writing to a RandomAccessFile, (2) writing to a file opened with O\_DIRECT and SYNC flags and (3) writing to a RAW partition. The RandomAccessFile requires a byte array stored in Java heap to read and write to disk. The other two access modes operate on page-aligned native arrays. In order to support both, we use ByteBuffers throughout the entire logging module. The ByteBuffers used for the RandomAccessFile are allocated in Java heap which allows accessing the underlying byte array used to read/write to disk. The ByteBuffers for direct access are stored in native memory by using the method \texttt{allocateDirect}. The access to the underlying byte array in native memory is done in a Java Native Interface (JNI) module by accessing it directly by address. The address is determined by using the call \texttt{Buffer.class.getDeclaredField("address")}. To avoid calling the reflecting method every time the buffer is accessed in the JNI module, we determine the address once and store it alongside the reference of the ByteBuffer in a wrapper which is used throughout the logging module. A performance comparison between Direct-, HeapByteBuffers and arrays can be found in Section \ref{eval_buffers}.

Most of the ByteBuffers used to write to or read from disk are pooled to relief the garbage collection and speed-up the processing. The only two exceptions are the buffers used to write to primary log (length of all log entries to write to primary log differs significantly from write to write) and to flush the secondary log buffer if the new bulk to write exceeds the secondary log buffer size (e.g., 100 KB in secondary log buffer and 150 KB in write buffer). 

During fault-free execution the current position within a log/segment and the length of a log/segment is stored in RAM (for performance reasons). For the recovery of a failed master, the information, stored on backup servers, is used as well. However, in case of a power failure the lengths and positions (irrelevant for the recovery) are unavailable. We cannot store the lengths on disk because this is too slow. Instead, every log's file is initialized with zeros and every write access to primary and secondary logs is followed by a $0$ to mark the end. Whenever a log entry is read which starts with a $0$, we know that the end of the segment/log is reached as the type field of a log entry header cannot be $0$. And we do not have to mark the end of the version logs as the files are truncated after every write.

\subsubsection{RandomAccessFile}
\label{raf}
The RandomAccessFile is probably the easiest and most comfortable way for random writes and reads to/from a file in Java. The RandomAccessFile is based on Java's \texttt{FileInput-} and \texttt{FileOutputStream} which use the read and write function of the operating system. In Linux all write and read accesses are buffered by the page cache (if the file was not opened with O\_DIRECT flag), i.e., when writing to disk the buffer is first written to the page cache and eventually to disk (may be cached on disk as well). We discuss the dis-/advantages of the page cache later.

We create one file for every log (primary, secondary and version logs) in the file system (e.g., ext4). The files are opened in read-write mode ("rw"). Before every read/write access we seek to the position in file. The offset in the byte array can be passed to the read/write method.

\subsubsection{Direct Access}
\label{direct_access}

\begin{figure}[!t]
	\centering
	\includegraphics[width=3.5in]{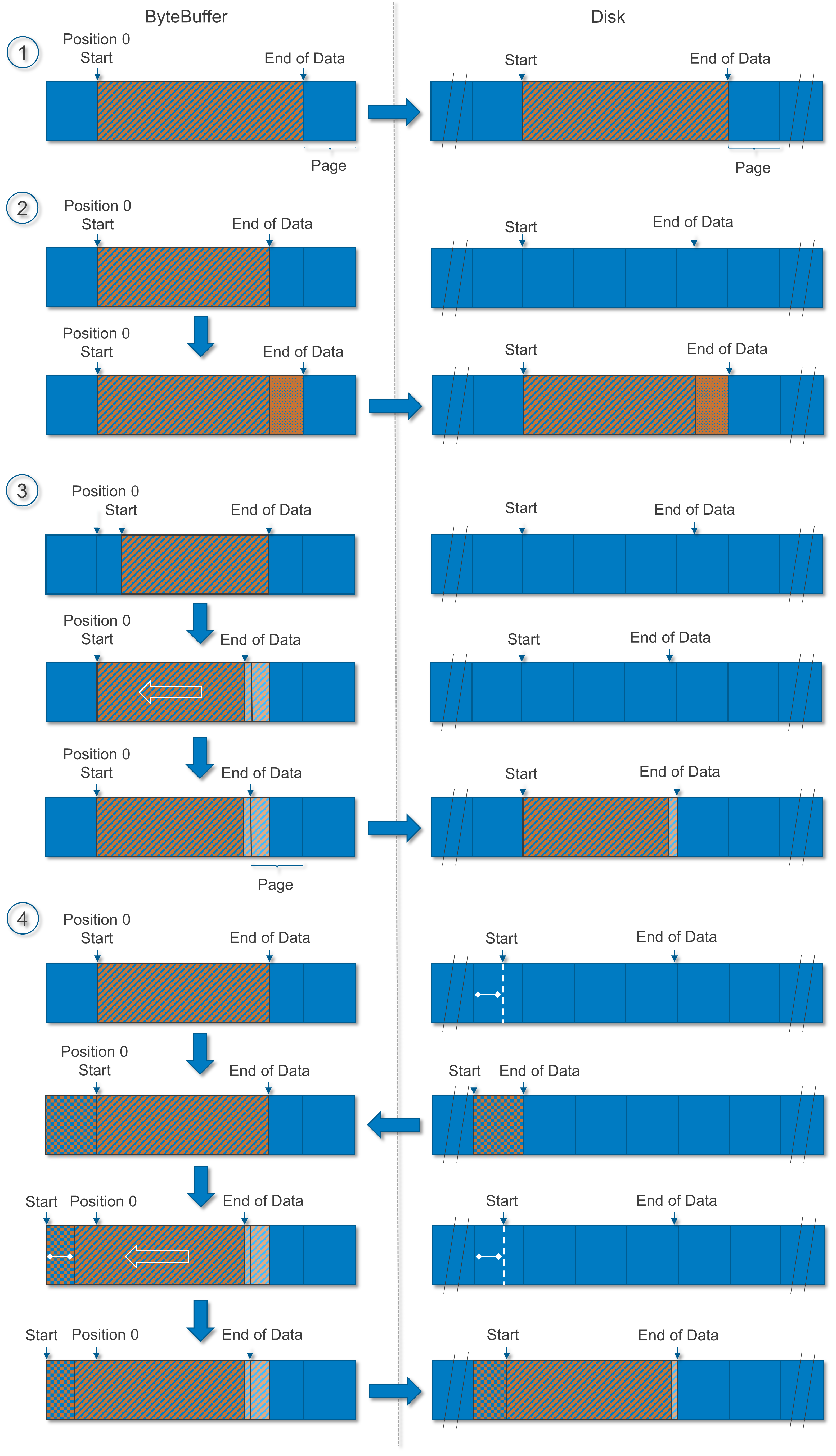}
	\caption{Buffer and File Alignment}
	\label{alignment_fig}
	\vspace{-20pt}
\end{figure}

To directly access a file, we have to use Linux functions, which cannot be accessed in Java. We use JNI to integrate a $C$ program which handles the low-level access to files. Files are opened with \texttt{open}, read with \texttt{pread} and written with \texttt{pwrite}. The Linux kernel functions \texttt{pread} and \texttt{pwrite} have the same behavior as \texttt{read} and \texttt{write} but do not change the file pointer. The buffers' addresses are passed as longs and the file IDs as ints and are both managed in the Java part. Accessing files directly, without page-cache, requires the files to be opened with the O\_DIRECT flag. We open logs with the following flags:
\begin{itemize}
	\item O\_CREAT: create the file if it not already exists
	\item O\_RDWR: this file is going to be read and written
	\item O\_DSYNC: write accesses return after the data was written (might be in disk cache). Metadata (e.g., timestamps) might not be updated yet
	\item O\_DIRECT: this file is accessed directly without page-cache
\end{itemize}
After opening the file, we write zeros to the file by calling \texttt{fallocate} which also reserves the memory for the entire log. If \texttt{fallocate} is not available (e.g., for ext2), we use \texttt{ftruncate}, which is noticeably slower. It is mandatory to reserve the memory for the entire log when creating/opening the file as write accesses that require enlarging the file and appending the data might use the page cache again.

The most important difference when accessing files directly is that every read and write access must be page- and block-aligned (typically, the page size is a multiple of the block-size). This means, both the position in file and in buffer must be page-aligned, as well as the end of the read/write access. In Sections \ref{primary_log} to \ref{writing_to_logs}, we described the different disk accesses. In this section, we discuss the impact on the buffer and file position and how to handle the accesses correctly. 

There are only two read access patterns: (1) read an entire segment from a secondary log and (2) read an entire version log. In both cases, both the file position (either $0$ or a multiple of the segment size which is a multiple of the page size) as well as the buffer position (always $0$) is page-aligned. Thus, for read accesses we do not have to consider the alignment. The function \texttt{pread} might return before all data was read. Therefore, all read accesses are executed in a loop which breaks if all data have been read or the end of the file has been reached. 

We discuss the write access patterns separately:

\textbf{Compacting a version log:} Prior to the reorganization of a secondary log, the version log is read-in, compacted and written back to disk. In this case, the buffer position is page-aligned because the entire buffer is written (from buffer position $0$). The file position is page-aligned as well because we write to file position $0$ (see Figure \ref{alignment_fig} situation 1). If the end is not page-aligned (same for buffer and file), we write the entire last page (see Figure \ref{alignment_fig} situation 2). This requires the buffer being larger than the data. We discuss the buffer allocation which considers writing over the data boundaries at the end of this section. After the write access, the version log is truncated with \texttt{ftruncate}.

\textbf{Writing to a new segment in secondary log:} When writing to a new segment in a secondary log, the file position is page-aligned as the position is a multiple of the segment size (which is a multiple of the page size). If the buffer position is $0$, the buffer can be written to the corresponding secondary log like the version log (see Figure \ref{alignment_fig} situation 1 and 2). Secondary logs are accessed segment-wise and within a segment, data is always appended. The end of a segment is also the end of a page. Therefore, writing to a segment does not affect the following segment, even if the last page is filled with invalid data to page-align the write access. The position in buffer is not always $0$. Sometimes, a segment is filled up and the rest of the buffer is written to a new segment. In this situation, if the position is not a multiple of the page size, we move the complete data to the beginning of the page of the first byte (with \texttt{memmove}) and write to the file from this position in the buffer (see Figure \ref{alignment_fig} situation 3). The end is handled as before.

\textbf{Flushing a version buffer to its version log:} Whenever the version buffer is full or the threshold is reached, the entire version buffer is flushed to the end of the version log. In this case, the buffer is aligned, but the file position most likely is not as we append at the end of the version log and a version number has a size of $13$ bytes. Therefore, when writing to the version log, all bytes from the last written page of the file have to be read and put in front of the data in the ByteBuffer to write (see Figure \ref{alignment_fig} situation 4). Then, the data is moved to the offset in file (start position $\%$ page size) within the ByteBuffer. All buffers have one additional free page in front of the start position for this situation (see the end of this section). Again, if the end position is not aligned, the last page must be written entirely and the file is truncated afterwards.

\textbf{Writing to primary log:} This is mostly the same situation as flushing to a version log. The buffer position is $0$ and the file position arbitrary (see Figure \ref{alignment_fig} situation 4). However, the file is not truncated afterwards.

\textbf{Appending to a segment in secondary log:} When appending to a segment, both the position in buffer as the position in file is most likely not page-aligned. This is a combination of situation 3 and 4 in Figure \ref{alignment_fig}. However, as all bytes to write in the buffer have to be moved anyway (in situation 4), the write access is handled like described in situation 4 of Figure \ref{alignment_fig}.

\textbf{Freeing a segment in secondary log:} If all log entries of a segment are invalid during the reorganization, the segment must be marked as free. This is done by writing a $0$ to the beginning of the segment. As it is not possible to write a single byte, we write an entire page filled with zeros.

All write accesses are executed in a loop because \texttt{pwrite} might return before all bytes have been written.

\textbf{Buffer Allocation:} The write accesses, as described, do not need allocations or to copy data to other buffers. In some cases, data is moved within the write buffer and data is read from file to the buffer beyond the boundaries of the buffer. This must be considered for the buffer allocation as well as the page alignment of the buffer. All buffers used for writing to disk are allocated in a wrapper class which stores the buffer's address and the ByteBuffer's reference. The ByteBuffers are allocated with \texttt{ByteBuffer.allocateDirect()} and the byte order is set to little endian. A ByteBuffer created with aforementioned method in most cases is not page-aligned. Hence, we create a ByteBuffer which is exactly one page larger than required. Then, we set the position to $address$ $\%$ $page$ $size$ and the limit to $position$ $+$ $requested$ $length$. Finally, we slice the ByteBuffer to create a second ByteBuffer instance which refers to the same byte array in native memory but with the position and limit of the first instance as beginning and capacity.

The ByteBuffers must not only be page-aligned, but also have one free page in front of it and the last page must be allocated entirely, as well, if the end is not page-aligned. Therefore, we add another page and the overlapping bytes to the size of the ByteBuffer. Furthermore, the address and the beginning of the sliced buffer is set to the page-aligned offset plus one page. The additional memory is not a problem because the buffers are rather large. For all buffers used in the logging module, we need less than 160 KB additional memory (see Section \ref{memory_consumption}).
		
\subsubsection{Raw Access}
\label{raw_access}
The RAW access is based on O\_DIRECT and shares the read and write functions. The difference is that the direct access method uses files provided by the file system whereas the RAW access method accesses a raw partition instead. This way, we can reduce the overhead of a file system like timestamps, many indirections and journaling. Furthermore, we can optimize the structure for the only purpose of storing logs (appends, no deletes).

\begin{figure}[!t]
	\centering
	\includegraphics[width=3.5in]{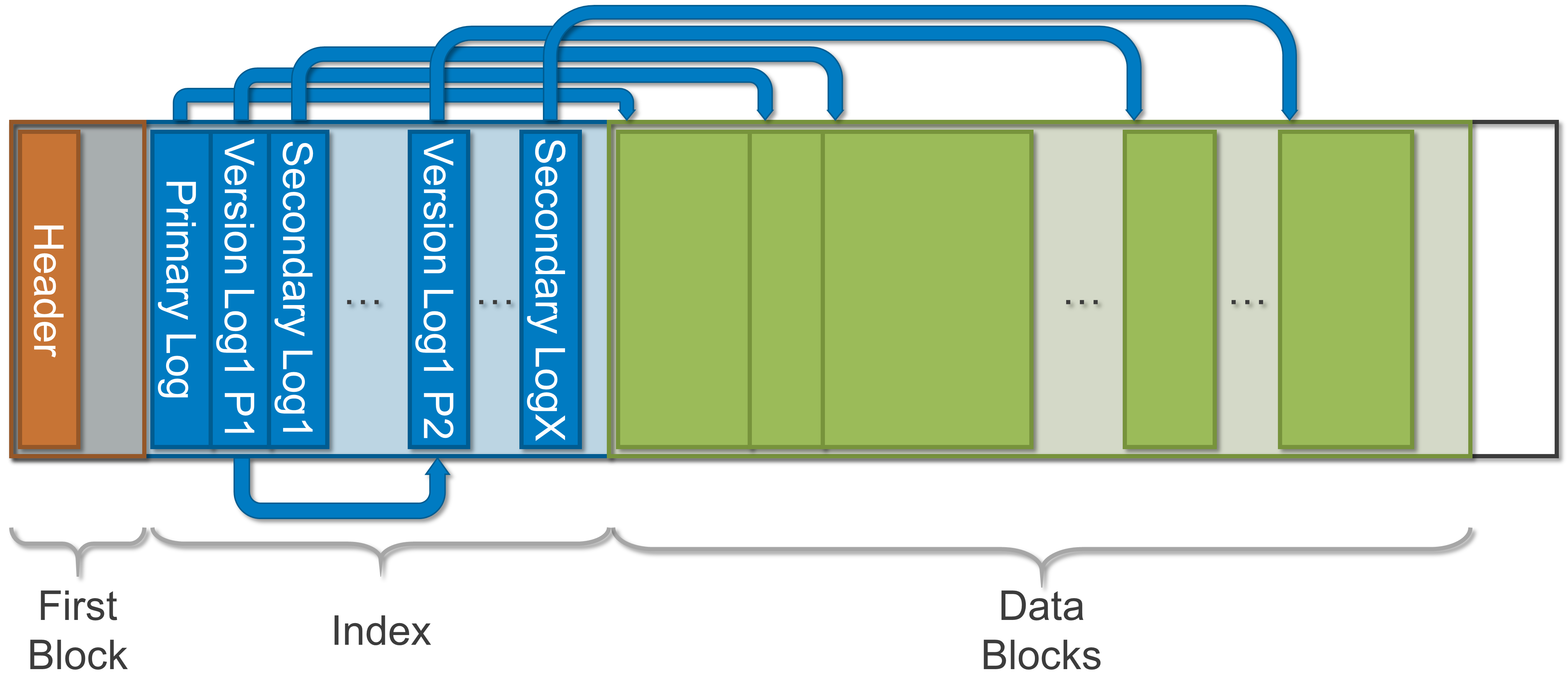}
	\caption{Structure of the RAW partition}
	\label{raw_fig}
	\vspace{-20pt}
\end{figure}

The raw access requires structuring the partition to access different logs. We divide the raw partition into three parts (see Figure \ref{raw_fig}): a start block with a partition header (for identifying a partition after failure), an index for finding logs and a data block which stores the logs.

The index has a tabular form. Every row is a 64-byte index entry containing an address within the partition pointing to a log, the log's name, type (primary, secondary or version log) and size. One difficulty is that the version logs can grow and shrink (the other logs have a fix size). Therefore, an index can point to another index entry for indexing the next part of a log. Initially, a version log is created by appending one index entry and one 16 MB data block. If the version log grows beyond 16 MB, another index entry and 16 MB data block are created. To find the second part of the version log, the first index entry stores the address of the second index entry.

The index block is cached in RAM for fast indexing. When the index was changed, the manipulated entries are written to disk (page-wise). Obviously, write accesses to the index block must be synchronized. The read and write accesses to the logs must not.

\subsubsection{Comparison of the Access Methods}
The default access method in DXRAM is direct access. Compared to the RAW access, it is more versatile as logs can be stored on every partition with a file system. RAW requires a dedicated raw partition which cannot not be provided on every server. Furthermore, the written logs can be analyzed with other tools when stored as a file. Table \ref{odirect_comparison} shows the dis-/advantages of using O\_DIRECT in comparison with using the page cache. Generally, the disadvantages outweigh the advantages. However, DXRAM's demands are uncommon. First of all, write and read accesses are quite large. Reads are usually at least 8 MB, writes are as large as possible without impairing the reliability. During a typical load phase, write accesses are between 7 and 8 MB on average. Furthermore, a exemplary backup server stores seven times the amount it stores in RAM and most of the RAM is occupied by in-memory objects which strictly limits the size of the page cache. Thus, caching is not very effective for DXRAM's logging. On the contrary, the disadvantages of utilizing the page cache weigh much more. The double buffering is not efficient and the page cache cannot be restricted in size and may grow rapidly. Whenever the application needs more memory and all memory is in use, the page cache must be flushed to disk, which can take a while. Furthermore, if the page cache contains many dirty pages, flushing the cache can pause the entire system for several seconds (amount of dirty pages can be configured). Another problem is the reliability. When a server crashes all dirty pages are lost. Since the DXRAM does not know when the write access is flushed to disk, DXRAM and its applications might be in an inconsistent state when rebooting. The RandomAccessFile allows synchronous disk access, but this is slow in comparison to the implemented access via O\_DIRECT as all data has to be copied to the page cache anyway.

\captionsetup{font=sc}
\begin{table}[!t]
	\centering
	\normalsize
	\caption{Dis-/Advantages of using O\_DIRECT}
	\begin{tabular}{|l|}
		\hline
		\textbf{Advantages:} \\ \hline
		Lower and predictable RAM consumption (no caching) \\ \hline
		Synchronous access without copying \\ \hline
		\textbf{Disadvantages:} \\ \hline
		More complex to use \\ \hline
		No performance benefits from caching \\ \hline
		Dependent on the underlying system \\ \hline
		No asynchronous write access supported \\ \hline
	\end{tabular}
	\label{odirect_comparison}
\end{table}

\section{Logging in DXRAM - Metadata Overhead}
\label{memory_consumption}

\captionsetup{font=sc}
\begin{table*}[!t]
	\centering
	\normalsize
	\caption{Memory consumption for N backup zones}
	\begin{tabular}{|l|l|l|}
		\hline
		Data Structure & Quantity & Aggregated Memory Consumption \\ \hline
		Write Buffer &  $1$ & $32$ MB \\ \hline
		Secondary Log Buffers & N & N$* 128$ KB \\ \hline
		Version Buffers & N & N$* 3$ MB \\ \hline
		Pooled Buffers for Secondary Logs & $8*8$ MB$+32*1$ MB$+64*0.5$ MB & $128$ MB \\ \hline
		Pooled Read/Write Buffers for Version Logs & $2$ & $\sim 50$ MB \\ \hline
		Pooled Buffer for Reorganization & $1$ & $8$ MB \\ \hline
		Pooled Buffers for Recovery & $5$ & $40$ MB \\ \hline
		Range Sizes Hash Table & $1$ & $28$ KB \\ \hline
		Range Buffers Hash Table & $1$ & $44$ KB \\ \hline
		Version Hash Table for Reorganization & $1$ & $\sim 45$ MB \\ \hline
	\end{tabular}
	\label{data_structures_table}
\end{table*}

Table \ref{data_structures_table} shows the memory usage of all data structure used in DXRAM's backup system (logging, reorganization and recovery). When a backup server stores 1000 backup zones with an average chunk size of 64 bytes, resulting in around 1 TB on disk, the RAM usage would be 3.4 GB, which is around $\frac{1}{300}$ of the disk usage. All given values are optimized for performance, i.e., if the load is very high and every update belongs to the same backup zone, the performance would be optimal. The best way to reduce the memory usage is to shrink the version buffers. With 1 MB version buffers, DXRAM needs 1.4 GB or $<\frac{1}{700}$ of the disk space and the performance would be untouched for most situations (assuming that the access distribution is not extreme).

\section{Related Work on Segment Selection}
\label{related_work_segments}
In \cite{LFS}, Rosenblum et al. presented a file system which is based on a log structure, i.e., file updates are appended to a log instead of updating in-place. This allows aggregating of write accesses in RAM in order to efficiently utilize the disk by writing large batches. For given workloads the log-structured file system (LFS) utilizes the disk an order of magnitude more efficiently than an update-in-place file system. The work was inspired by write-ahead logs of databases and generational garbage collection of type-safe languages which also need to clean-up in order to reclaim space by removing invalid/outdated objects.

While not being the first developing a LFS, Rosenblum et al. contributed by analyzing workloads to find an efficient reorganization scheme. A fast reorganization is important to keep a constant write throughput (provide enough free space for writes) and to allow a fast crash recovery (less invalid/outdated objects to process). In \cite{LFS}, a log is subdivided into 8 MB segments. The reorganization selects a segment, reorganizes it and proceeds with another segment. Important for the efficiency of the reorganization is the segment selection. Optimally, the segment with most invalid/outdated data is selected for the cleaning. Rosenblum et al. stated the assumption that "the older the data in a segment the longer it is likely to remain unchanged" \cite{LFS}. This leads to the following cost benefit formula, which did well in the evaluation:

\begin{equation}
\begin{split}
& \frac{benefit}{cost} = \frac{free\ space\ generated * age\ of\ data}{cost} \\
& = \frac{(1 - u) * age}{1 + u}
\end{split}
\end{equation}

$u$ is the utilization of the segment which is the fraction of data still live, $age$ is the age of the youngest block within a segment.

Seltzer et al. did a more thorough performance analysis on log-structured file systems showing the high performance impact of the cleaning (more than 34 \% degradation if cleaning is necessary) \cite{LFSPerf}. In \cite{ramcloudLogging}, Ousterhout et al. applied the ideas of a LFS for the in-memory key-value store RAMCloud. RAMCloud uses a log for storing in-memory objects and replicates the objects segment-wise to remote disks. Furthermore, they present a two-level cleaning approach which is a combination of in-memory reorganization and disk compactification. In RAMCloud, all complexity resides on the masters, storing the objects in RAM. The backup servers are used for the plain writing to disk. Therefore, the backup servers cannot execute the reorganization (they miss information like the current version numbers), but they can compact logs from time to time. To avoid recovering already deleted objects, RAMCloud's masters write \textit{tombstones} (difficult to remove) to the logs. DXRAM, on the other hand, stores the in-memory objects with a tailored memory management in RAM and replicates the objects to backup servers as soon as they are written. The backup servers perform the version control and reorganization of all of its stored objects without communicating with masters. DXRAM avoids tombstones as backup servers can identify deleted objects through the version control.

In \cite{ramcloudCostBenefit}, Rumble et al. modified the cost benefit formula for RAMCloud:

\begin{equation}
\begin{split}
& \frac{benefit}{cost} = \frac{(1 - u) * segmentAge}{u}
\end{split}
\end{equation}

The first difference, regarding the denominator (from $1 + u$ to $u$), considers that RAMCloud does not have to read the segment from disk prior to the reorganization as all segments are stored in RAM on masters. The second change concerns the age which is not the age of the youngest block, anymore, but the average age of all objects in a segment. The latter avoids the unnecessary reorganization of segments which have a high utilization and store mostly old objects but one or a few new objects.

\section{Segment Selection in DXRAM}
\label{segments}
In DXRAM, the segment selection requires two steps because DXRAM does not store one log with all objects of a master but one secondary log per backup zone (a master can have hundreds of backup zones). The first step is selecting a secondary log to be reorganized and the second step is selecting segments of this secondary log. The secondary log to reorganize is chosen by its size: the largest log has the most invalid data as backup zones are identical in size (assuming there is no fragmentation). For selecting a segment, we cannot adopt RAMCloud's approach for DXRAM because instead of an in-memory log on the masters DXRAM uses an in-place memory management on the masters and a separated log-structure on backups. Therefore, the log entries on the backups are never read individually but as whole segments. This allows us to spare storing the locations of log entries within a log saving a lot of memory on masteres (e.g., for one billion 64-byte chunks and three replicas: $> 30$ GB per master are saved). But, without the location of invalid log entries, it is not possible to determine the fraction of live data of a segment. Obviously, searching for invalid versions to update the segments' utilizations is not an option.

In the following we use a different definition of the term \textit{utilization} (in comparison to \cite{LFS} and \cite{ramcloudCostBenefit}). We define the utilization $u$ as the plain filling degree of a segment (live, outdated and deleted chunks).

\subsubsection{Basic Approach}
\label{basic_approach}
A secondary log is never reorganized as a whole but incrementally by reorganizing single segments (default: 8 MB). Similar to the secondary log selection, the segment selection tries to find the segment with the most outdated data. In the basic approach, we calculate a segment's age based on its creation and last reorganization and select an old segment with high utilization for cleaning ($max(age * utilization)$). We think this is a good metric as there is a higher probability of finding outdated objects in segments that are large and have not been reorganized for a longer period of time. Additionally, this approach is very simple to implement and comes at no cost as all required metadata is already available.

\subsubsection{Advanced Approach}
The advanced approach tries to improve the log selection, i.e., selected segments contain the most outdated data, by including additional or more precise indicators. The decision making must consider the following constraints: (1) neither the exact location of an object nor the segment an object is stored in is available because (2) object specific information cannot be stored in RAM due to the memory consumption being too high. Therefore, (3) the utilization as described in \cite{LFS} and \cite{ramcloudCostBenefit} representing the fraction of valid data is not available, either, because maintaining the information would require the location of previous versions.

\textbf{Utilization: } The utilization (filling degree) is a good indicator for the segment selection if all chunks are accessed evenly because the segment with highest utilization would, on average, have the most invalid data. But, segments with much cold, long-living data are chosen repeatedly blocking the reorganization of segments with (more) invalid data. Therefore, the utilization alone is not a good indicator in every scenario.

\textbf{Age: } In Section \ref{basic_approach}, we defined the age of a segment as the time since the last reorganization or creation. While this approach is easy to implement, the validity of the age is highly limited as the age of long-living objects is not covered (the time is reset regardless of whether much data was discarded or not) and a freshly reorganized segment is not necessarily used next to add new objects. The least recently reorganized segment might even store the same still valid objects. In this section, we discuss an approach to determine a segment's age based on the age of all containing objects.

Rosenblum et al. state that "the older the data in a segment the longer it is likely to remain unchanged" \cite{LFS}. This claim cannot be transferred to DXRAM because it is based on the assumption that updated and deleted data is marked invalid in the segment headers and the age is determined for the valid data, only. In DXRAM, instead, invalid data is exclusively detected and discarded during the reorganization, i.e., a segment's age is the average age of all, valid and invalid, objects. Without the implication regarding the validity of an object, an object's age has to be interpreted differently: typically, older objects are more likely to be deleted or updated. Thus, a segment with more old objects might be the better choice for the reorganization. But, often objects can be split into the two categories: hot and cold data. Cold data consists of long-living objects that are unlikely to be replaced/removed. Therefore, a segment with very old objects might not be the best choice. Altogether, the age of an object is an important indicator for the validity of an object.

\textbf{Average age per entry: } To get a more accurate representation of a segment's age, we store a 4-byte timestamp in the log entry header of all objects (stored in front of the object on disk). An empty segment has the age $0$. After the reorganization of a segment its age is defined by:

\begin{equation}
\begin{split}
& a_{seg_i} = \frac{\sum\limits_{j=0}^n t - t_{c_j} | c_j\ valid}{m}
\end{split}
\end{equation}

$n$ is the number of objects in segment $i$ and $m$ the number of valid objects ($c$ for chunk). As every object ages between two reorganizations, when selecting a segment, we adjust the average age of a segment by adding the time since its last reorganization.

\begin{equation}
\begin{split}
& a'_{seg_i} = a_{seg_i} + (t - t_{reg})
\end{split}
\end{equation}

Assuming we add a new object ($c_x$) at the same time the reorganization is executed, then the age is modified in the following way:

\begin{equation}
\begin{split}
& a_{seg_i} = a_{seg_i} + \frac{t - t_{c_x} - a_{seg_i}}{n + 1} \\
& = a_{seg_i} - \frac{a_{seg_i}}{n + 1}
\end{split}
\end{equation}

When adding an object after the reorganization, we have to consider the time since the last reorganization to avoid increasing the age too much during the segment selection as a segment's age is based on all objects' ages at the time of the last reorganization. The object did not exist at this time. Therefore, we have to subtract the time difference.

\begin{equation}
\begin{split}
& a_{seg_i} = a_{seg_i} - \frac{a_{seg_i} + (t - t_{reg})}{n + 1}
\end{split}
\end{equation}

\textbf{Average age per byte: } Objects might differ significantly in size. To avoid missing segments with large old and invalid objects (to be discarded) and many small young objects (decreasing the age), we calculate the age per byte and not per chunk. Furthermore, we exclude every object which is older than a predefined threshold and still valid (hot-to-cold transformation).

\begin{equation}
\begin{split}
& a_{seg_i} = \frac{\sum\limits_{j=0}^n (t - t_{c_j}) * s_{c_j} | c_j\ valid\ and\ t - t_{c_j} < t_{max}}{\sum\limits_{j=0}^m s_{c_j}}
\end{split}
\end{equation}

\begin{equation}
\begin{split}
& a_{seg_i} = a_{seg_i} - \frac{(a_{seg_i} + (t - t_{reg})) * s_{c_x}}{u_{seg_i} * s + s_{c_x}}
\end{split}
\end{equation}
$s$ is the segment size (e.g., 8 MB) and $s_{c_j}$ the size of $chunk_j$. $u_{seg_i}$ is the utilization (filling degree) of the segment.

\textbf{Utilization \& Age: } The final segment selection is based equally on the utilization and age of a segment:

\begin{equation}
\begin{split}
& seg = i \in \{1, ..., l\} | max(u_{seg_1} * a'_{seg_1}, ..., u_{seg_l} * a'_{seg_l})
\end{split}
\end{equation}

$l$ is the number of segments of the secondary log, excluding segments which have not been used yet (at the end of the log).

\textbf{Timestamps: } The used 4-byte timestamps show the elapsed seconds since the secondary log creation. An overflow occurs after more than 68 years and affects the segment selection (wrong decisions) for a short time, only.

\section{Related Work on Copysets}
\label{related_work_copysets}
In \cite{copysets}, Cidon et al. present a replication scheme which, in comparison to random replication, significantly reduces the frequency of data loss events in exchange for a larger amount of lost data in case of a data loss event. The authors motivates that restoring the data after a data loss event has fixed costs regardless of the amount of lost data. Thus, losing a large amount of data seldom is more attractive for cluster operators than losing small amounts frequently. The probability for data loss is rather high with random replication in large clusters because, assuming the number of objects is high, every master most-likely stores replicas on every available backup server. Thus, a failure of $x$ backup servers, where $x$ is the number of replicas for every object, results in a data loss event. The basic idea in \cite{copysets}, is to limit the number of backup servers one master replicates its data to. The limited set of available backup servers for a master is called a \texttt{copyset}. Subsequently, data loss is possible, if a set of $x$ backup servers of one copyset crash, only. Assuming the number of backup servers per copyset $R$ (which is also the replication factor) is much lower than the total number of backup servers $N$, the probability will be much lower. The authors exemplify two scenarios to prove their statement: (1) in a 5000-node RAMCloud cluster, copyset replication reduces the data loss probability from 99.99\% to 0.15\%. (2) In a HDFS cluster with a workload from Facebook, the probability is reduced from 22.8\% to 0.78\%.

The copysets are created by permuting the $N$ backup servers and assigning $R$ consecutive backup servers from the permutations to a copyset. The number of permutations $P$ is determined by:
\begin{equation}
\begin{split}
& P = ceil(S / (R - 1))
\end{split}
\end{equation}
$R$ is the number of servers per copyset and $S$ is the scatter width that defines the number of backup servers one master replicates its data to. If the scatter width is higher, more permutations are generated which results in one server being in more copysets.

Example: $N = 6, R = 3, S = 4$. The number of permutations is $2$ then and two permutations could be $5, 1, 3, 4, 6, 2$ and $6, 1, 2, 3, 4, 5$. Hence, the copysets are $\{5, 1, 3\}$, $\{4, 6, 2\}$, $\{6, 1, 2\}$ and $\{3, 4, 5\}$. To determine the backup servers for an object, the first backup server is chosen randomly. Afterwards, one copyset that includes the randomly chosen backup server is selected and the other servers in the copyset are assigned as additional backup servers. In the example above, if backup server $3$ is chosen randomly, the other backup servers are either $5$ and $1$ or $4$ and $5$ (scatter width in example is $3$, only, but with $N >> S$ a scatter width of $4$ is very likely). Every primary backups' files are distributed to the same set of backups ($1$, $4$ and $5$). Only, if all nodes from one copyset fail simultaneously data loss occurs. The scatter width is important for the recovery as it defines the number of servers which can recover a failed server in parallel.

In the next two sections, based on \ref{copysets}, we further discuss copysets on two systems: HDFS and RAMCloud \cite{copysets}. In Section \ref{copysets_in_dxram}, we use the same example to present the copysets implementation of DXRAM.

\subsection{Copysets in HDFS}
The Apache Hadoop Distributed File System (HDFS) is based on the Google File System (GFS) \cite{GFS} but has significant differences regarding the node allocation and chunk location management. HDFS is part of the open source programming MapReduce framework Hadoop. HDFS was designed to run on commodity hardware and applications with big data sets \cite{hdfs}. It has a master-slave architecture with a single master, called NameNode, and multiple slaves. The NameNode is responsible for the metadata and the slaves, also called DataNodes, for storing the data. For fault-tolerance reasons the data is replicated to other DataNodes, in 64 MB blocks.

\begin{figure}[!t]
	\centering
	\includegraphics[width=3in]{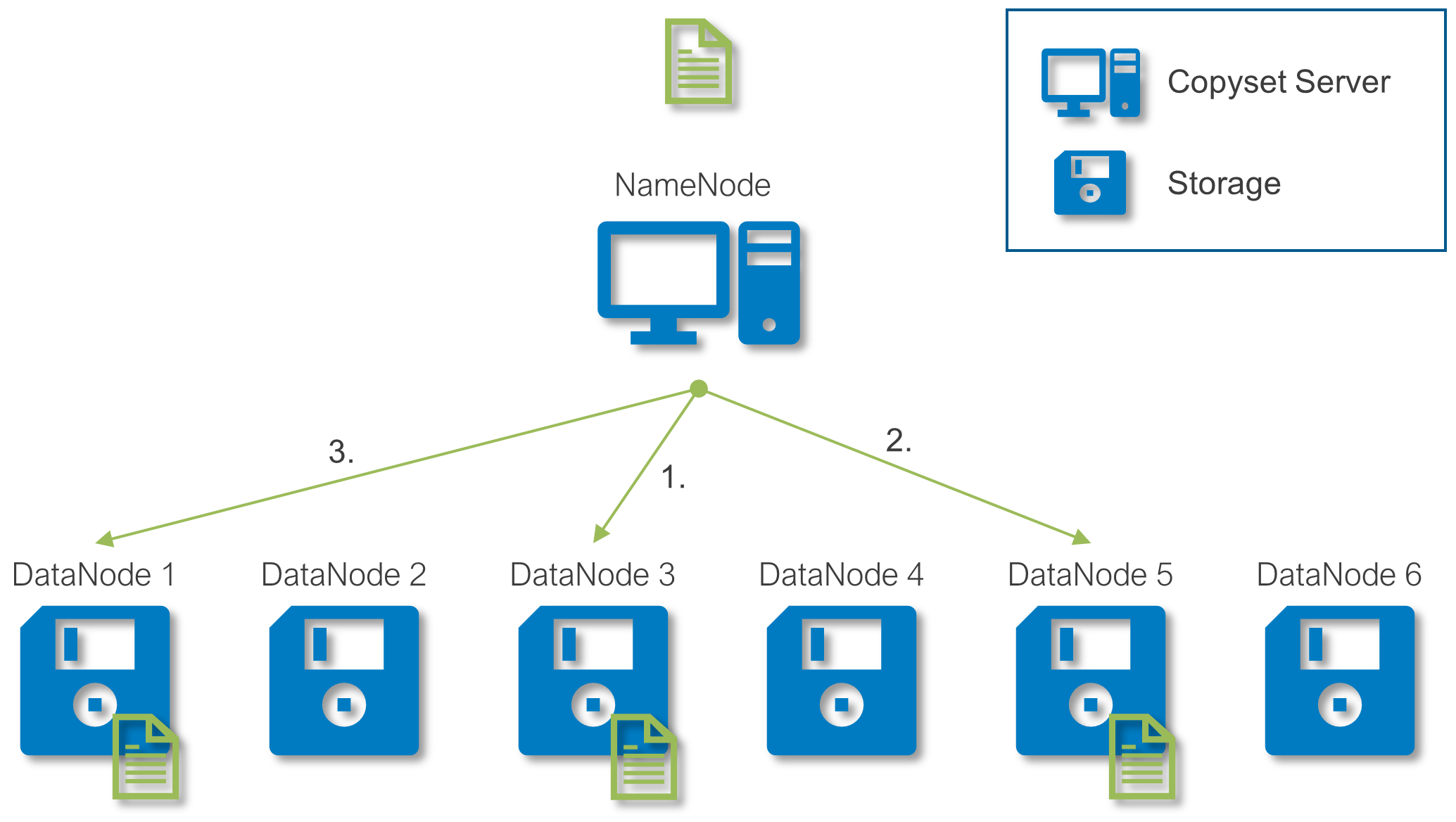}
	\caption{Copyset Determination in HDFS}
	\label{copysets_hdfs_fig}
	\vspace{-20pt}
\end{figure}

When using copysets for HDFS, the NameNode creates the copysets at system startup like described in the previous section. Every time a new file has to be stored, the NameNode chooses the first location randomly and then $R - 1$ DataNodes belonging to one of the copysets the first DataNode is in (Figure \ref{copysets_hdfs_fig}).

When a new DataNode is added to the system, the NameNode generates $S / (R - 1)$ new copysets which contain the new server. When a server crashes, it is replaced randomly in all copysets. In the example from the previous section, if DataNode $2$ crashes, the copyset could be modified in the following way:
$\{5, 1, 3\}$ $\{4, 6,$ \sout{2} $3\}$ $\{6, 1,$ \sout{2} $5\}$ $\{3, 4, 5\}$.

Subsequently, if another DataNode with ID $7$ is added, two additional copysets would be generated, for example:
$\{7, 1, 4\}$ and $\{3, 7, 5\}$.

\subsection{Copysets in RAMCloud}
RAMCloud is described in Sections \ref{related_work_logging} and \ref{related_work_segments}.

\begin{figure}[!t]
	\centering
	\includegraphics[width=3in]{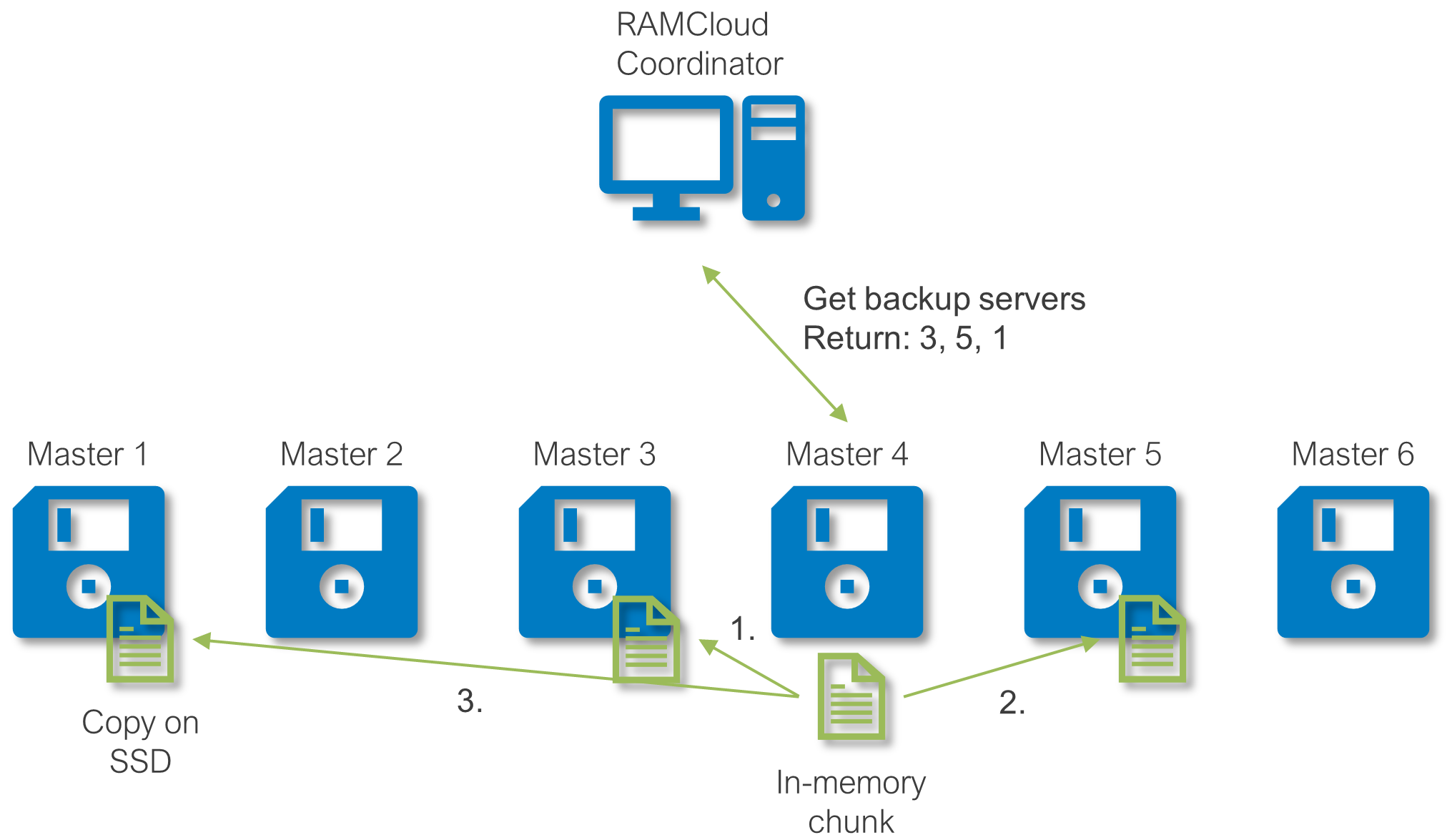}
	\caption{Copyset Determination in RAMCloud}
	\label{copysets_ramcloud_fig}
	\vspace{-20pt}
\end{figure}

In RAMCloud copysets are created on the coordinator. Whenever a master creates a new in-memory object, it queries a set of backup servers from the coordinator (Figure \ref{copysets_ramcloud_fig}). The first backup server is chosen randomly, the others belong to a copyset containing the first backup server. Every primary backups' objects are distributed to the same set of backup servers. But, every masters’ in-memory chunks are scattered across the cluster. Only, if all nodes from one copyset and the master fail simultaneously data loss occurs. The scatter width is \textbf{not} important for recovery as every master replicates its chunks to many copysets. Therefore, it is always $S = R - 1 = 2$ and the recovery time is nearly unaffected (1.1s instead of 0.73s \cite{copysets}).

In HDFS for every new DataNode $S / (R - 1)$ copysets are added. In RAMCloud, one has $S / (R - 1) = 1$ and instead of creating one new copyset directly, a new copyset is created when three new servers joined (all three servers are in the new copyset). When a server fails, it is replaced randomly like in HDFS.

\section{Copyset Replication in DXRAM}
\label{copysets}
In this section, we describe the most relevant aspects of DXRAM's backup zones in Section \ref{backup_zones}, followed by the copyset implementation of DXRAM in Section \ref{copysets_in_dxram}.

\subsection{Backup Zones}
\label{backup_zones}
In order to enable a fast parallel recovery, in DXRAM, the chunks of one server are partitioned into several backup zones (with a size of 256 MB) which are scattered across potentially many backup servers (e.g., a 64 GB server assigned with 256 different backup servers). Every server determines its own backup zones and informs its associated superpeer on each backup zone creation. This approach avoids global coordination regarding backup zone selection between servers. We use a replication factor of three by default but it is configurable.

Each backup zone is identified by a zone ID (ZID). The ZID alone is not globally unique but it is in combination with the creator's node ID derived from the context. A new backup zone is created whenever a chunk does not fit into any existing backup zone. If chunks were deleted, a backup zone will be gradually refilled with new chunks. Furthermore, chunks with reused CIDs are stored in the same backup zone as before, if possible, to minimize meta-data overhead. Three backup servers are assigned to each backup zone with a fixed replication ordering guaranteeing consistency. According to the ordering, the first backup server receives all backup requests first, the second afterwards and so on. Furthermore, backup requests are bundled whenever possible. If there are less than three servers currently available for backup (e.g., during startup), the next joining server will be used and receives all previously replicated chunks of this zone.

A server notifies its superpeer whenever a new backup zone was created or a backup server was changed. This results in a single message for every 256 MB (e.g., once after $3.5 \times 10^{6}$  64-byte chunks have been created) and a few messages per server failure (the failed backup server has to be replaced), only. To further reduce memory consumption on superpeers (resulting in just 10 bytes per backup zone in the best case), a superpeer does not store backup zone affiliations of chunks. This information is exclusively stored on the owner of a chunk as only this server must know the corresponding backup zone of its chunks for sending backup updates.

\subsection{Copysets in DXRAM}
\label{copysets_in_dxram}

\begin{figure}[!t]
	\centering
	\includegraphics[width=3in]{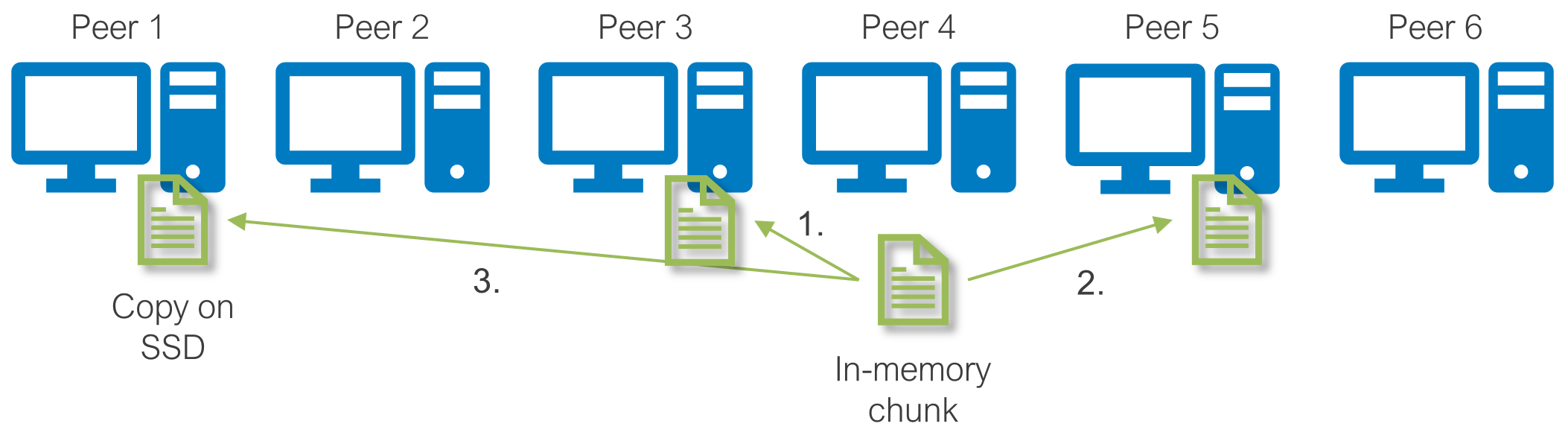}
	\caption{Copyset Determination in DXRAM}
	\label{copysets_dxram_fig}
\end{figure}

DXRAM does not have a coordinator, like the NameNode in HDFS or the coordinator in RAMCloud, but a set of servers responsible for the metadata (superpeers) and another set responsible for storing the data and backups (peers). Hence, we decided to create the copysets on every master independently but consistently by using the same input and algorithm to create copysets (no coordination needed). Consequently, every master also determines its own backup servers accordingly by choosing the primary backup server randomly and all other backup servers from one copyset containing the primary backup server (Figure \ref{copysets_dxram_fig}). Optionally, the primary backup server can be selected disjunctive and/or locality-aware. Another important difference is that DXRAM determines backup servers not for single chunks but for backup ranges containing many chunks (e.g., 256 MB). Therefore, the maximal number of copysets is smaller, if random replication is used. Still, with copyset replication the probability for data loss can be reduced.

For joining servers, we use the same strategy as RAMCloud: we wait for $R$ new servers to join and, then, create a new copyset containing all three servers. When a server crashed, it is replaced in all copysets. However, because of the decentralized copyset determination, we have to replace the failed server consistently on all masters. We do this, by using a seed which is based on the copyset (aggregated node IDs) for the pseudo random number generator.

The initial copyset determination is based on the nodes-file (a file used for startup which lists all servers participating) which is identical for all servers. Further un-/available servers are propagated by join and failure events which are distributed among superpeers first and to the peers afterwards. But, copysets can differ when servers are added because masters might detect the joining servers in different order. This case is rather unlikely but can occur from time to time. Therefore, the number of copysets (globally) can be higher than $N / R$ but still is a lot smaller (for $N >> S$) compared to random replication which is $N \choose R$ (e.g., with 512 backup servers $>=171$ for copyset replication and ${512 \choose 3} = 22,238,720$ for random replication; the number of combinations is limited by the number of backup zones in the system, for example 262,144).

In DXRAM, copyset replication can be combined with additional replication schemes like disjunctive first backup servers (to increase the throughput of the parallel recovery) and/or locality-awareness.

\section{Evaluation}
\label{evaluation}
In this section, we evaluate the byte array access methods as well as the disk access methods. Furthermore, we provide a thorough performance analysis on the logging and reorganization of DXRAM. The latter also includes a comparison of both presented segment selection strategies. 

All tests were executed on our non-virtualized cluster with 56 Gbit/s InfiniBand connection and servers with PCI-E nvme SSDs (400 GB Intel DC P3600 Series), 64 GB RAM, Intel Xeon E5-1650 CPU (six cores) and Ubuntu 16.04 with kernel 4.4.0-64.

\subsection{Byte Array Access}
\label{eval_buffers}
Log entries are almost always aggregated in larger buffers in the logging module. In order to find the best way to handle these buffers, we evaluated the different byte array access techniques provided by Java. We wrote a benchmark which writes to and reads from 8 MB buffers by using the access specific methods. The techniques are:

\begin{itemize}
	\item DirectByteBuffer BE: A ByteBuffer allocated outside the Java heap with big endianness.
	\item DirectByteBuffer LE: A ByteBuffer allocated outside the Java heap with little endianness (native order).
	\item HeapByteBuffer BE: A ByteBuffer allocated in the Java heap with big endianness (order of Java heap).
	\item HeapByteBuffer LE: A ByteBuffer allocated in the Java heap with little endianness.
	\item Array: A byte array in Java heap.
	\item Unsafe: A ByteBuffer allocated outside the Java heap with little endianness, accessed with methods provided by sun.misc.Unsafe.
\end{itemize}

Every buffer is filled first and then read entirely. We write/read a long value, followed by a short and three byte values, which is the access pattern of the version buffer and is also very similar to the access patterns of the primary and secondary log buffers. For representative results, we fill and read the buffers 1,000 times and ignore the first 100 iterations. In every iteration, we access another buffer. The buffers are allocated at the beginning of the benchmark to simulate the buffer pooling. For Java's Unsafe access, we also do boundary checks before every read and write access. Every test was executed five times.

\begin{figure}[!t]
	\vspace{-15pt}
	\centering
	\includegraphics[width=3.5in]{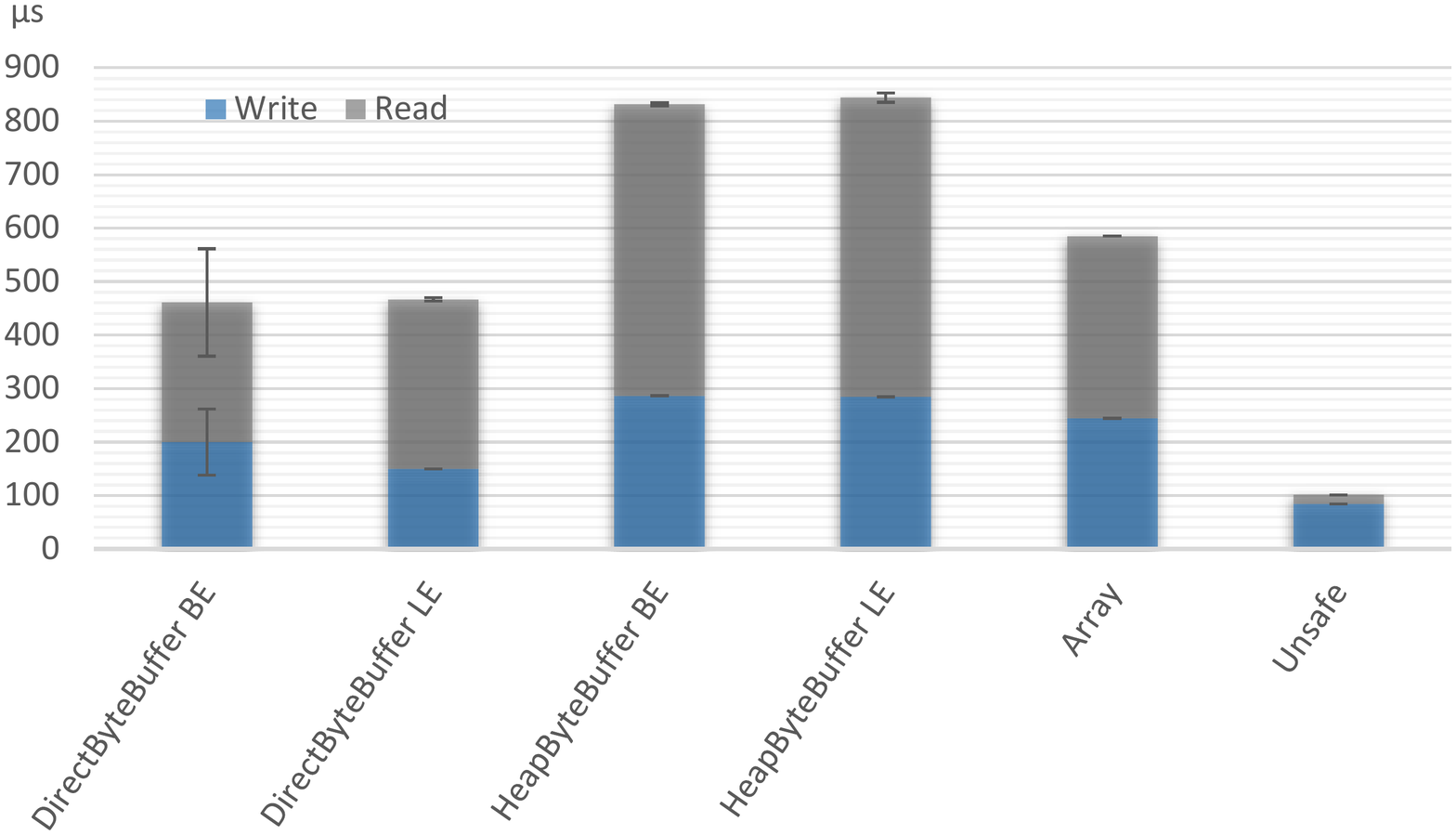}
	\vspace{-35pt}
	\caption{Evaluation of different byte array access methods. Writing and reading an 8 MB buffer (one to eight bytes per access) 900 times}
	\label{buffers_eval_fig}
\end{figure}

Figure \ref{buffers_eval_fig} shows the results of the presented benchmark. The benchmark runs are very consistent for all access methods but the DirectByteBuffer with big endianness. The native memory order on the used server is little endian. Therefore, the high variance can be explained by the byte swapping prior to every write access which is a rather CPU intense step. On the contrary, the Java heap is big endian. Thus, the variance of the HeapByteBuffer with little endianness is also higher than with big endianness. But, the difference is minor in this case. Nonetheless, using the endianness of the underlying memory for the ByteBuffer is advisable.

The DirectByteBuffer performs considerably better than the HeapByteBuffer and the heap array. Manipulating the data with Unsafe is even faster than with the DirectByteBuffer's methods. Subsequently, for the RandomAccessFile which needs a heap array for writing and reading, the fastest technique is the array itself. For O\_DIRECT and RAW access which requires the data to be off Java heap, Unsafe is the fastest choice. However, as the performance is relatively close, e.g., writing an entire 8 MB segment with longs, shorts and bytes is 120 ns slower with a DirectByteBuffer than Unsafe, and in order to reduce complexity (no wrapper, branching, dedicated serialization methods or boundary checks) and increase maintainability (debugging of segmentation faults is bothersome, future of Unsafe is unclear), we use ByteBuffers (DirectByteBuffer LE for O\_DIRECT/RAW and HeapByteBuffer BE for RandomAccessFile) for the logging module of DXRAM. Furthermore, one has to consider that this are the results of a micro benchmark and the real application's behavior is not identical.

\subsection{Logging and Reorganization}
\label{eval_logging}
In this section, we evaluate the logging and reorganization performance of DXRAM. First, we analyzed the maximum throughput of the SSD first. We used a SSD of the type Intel DC P3600 Series with a capacity of 400 GB. It provides a maximum throughput of 2.6 GB/s for read accesses and 1.7 GB/s for write accesses. The random I/O throughput of 4 KB chunks is capped at 450 MB/s for reads, 56 MB/s for writes and 160 MB/s for 70\% reads and 30\% writes. We measured the SSD performance with \texttt{dd} by writing 1,024 8 MB (default segment size) files (\texttt{/dev/zero}) with direct access.The results for 8 MB write accesses are significantly below the maximum throughput, showing 914 MB/s. With two processes writing concurrently, the throughput improves to 1,116 MB/s. With more processes the throughput is consistent (e.g., 1,170 MB/s with four processes). SSDs operate highly parallel and both the nvme driver (no I/O scheduler) and the Linux kernel (Multi-Queue Block IO Queueing \cite{multiqueue}) take advantage of that if read/write accesses are executed in parallel. DXRAM benefits from the parallelism by logging, reorganizing, recovering and reading/writing versions concurrently. When writing smaller chunks to disk, the throughput degrades, e.g., 477 MB/s for 8 KB chunks.

\subsubsection{Logging}
\begin{figure}[!t]
	\vspace{-15pt}
	\centering
	\includegraphics[width=3.5in]{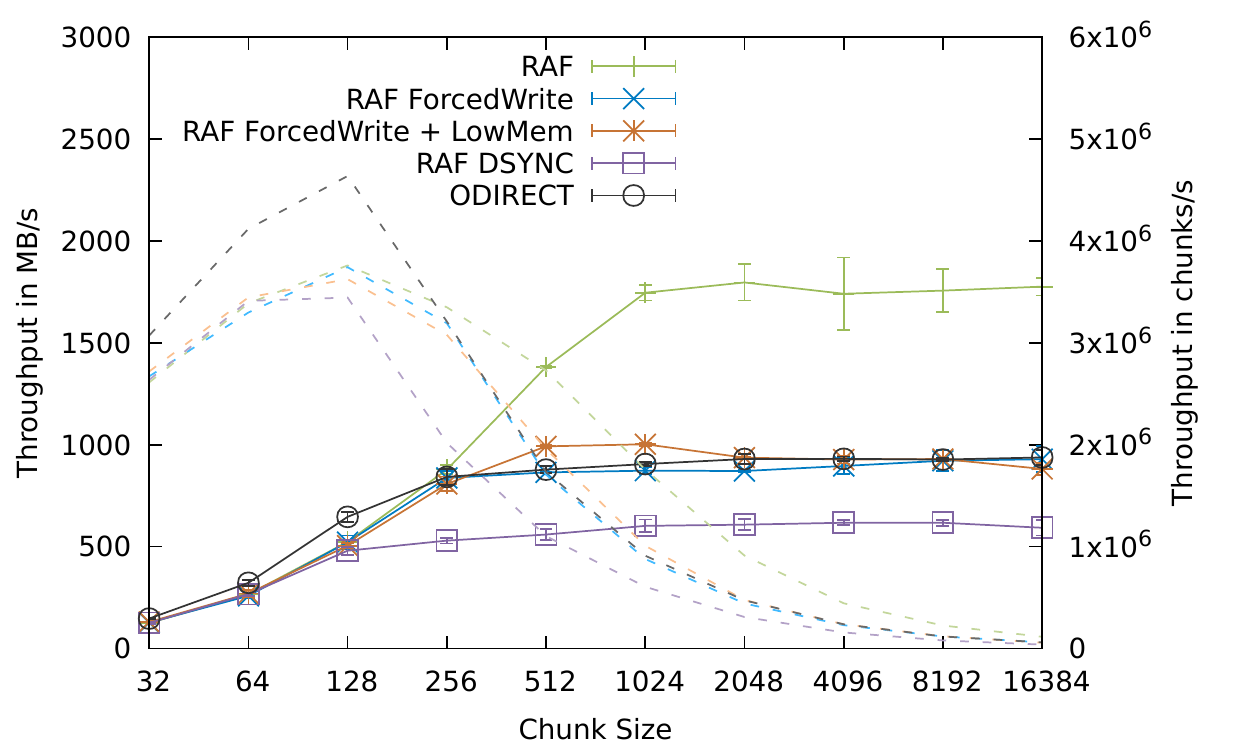}
	\caption{Evaluation of different disk access methods. Every chunk is written once, in sequential order. Solid lines: throughput in MB/s, dashed lines: throughput in chunks/s}
	\label{logging_eval_fig}
\end{figure}

Figure \ref{logging_eval_fig} shows the logging throughput (in MB/s and chunks/s) of the RandomAccessFile and O\_DIRECT for chunk sizes from 32 bytes to 16 KB. We evaluated the RandomAccessFile in four different configurations explained in the next paragraphs: (1) without limitations (RAF in Figure \ref{logging_eval_fig}), (2) with forced writes (RAF ForcedWrite), (3) with forced writes and with limited memory available (RAF ForcedWrite + LowMem) and (4) in synchronous mode (RAF DSYNC).

In order to provide resilience, DXRAM requires to store logged chunks persistently without much delay. However, In the default configuration, the OS caches many dirty pages in the page cache before eventually flushing the pages to disk. Therefore, when evaluating with forced writes, the OS is configured to flush dirty pages of the page cache reaching 8 MB, if possible, and immediately flush when 32 MB is dirty. The OS's flushing threads are also configured to flush more frequently (every 100 ms) than normal regardless of the two thresholds.

Even with a limited amount of dirty pages in the page cache, the page cache might grow critically for a memory-heavy application, i.e., the page cache occupies memory the application needs requiring to flush the page cache which might require many seconds. Therefore, we tested the logging with a limited amount of memory available by starting a program apriori which occupies most of the memory (92.5\%/59.2 GB).

Finally, the RandomAccessFile was opened in synchronous mode (\texttt{rwd}). In this mode, a write access returns after the data was written to disk. In contrary to \texttt{rws}, the file system's metadata (e.g., timestamps) might not have been updated. In all other test, the RandomAccessFile was opened with \texttt{rw} which is asynchronous.

The benchmark used to determine the logging throughput (and reorganization throughput in Section \ref{eval_reorg}), creates ten chunks (number configurable) and serializes them into a DirectByteBuffer. The buffer is passed to the logging component to be logged. For the next iteration the chunk IDs are incremented and the buffer is logged, again. This is repeated until the predefined number of chunks (e.g., 400,000,000 32-byte chunks) has been logged. The benchmark does not involve the network or any other components or services from DXRAM but the logging component. Every experiment is executed three times and old logs are removed and the SSD is trimmed (\texttt{fstrim}) between runs to get consistent results.

The RandomAccessFile without limitations (RAF) is the fastest disk access mode for chunks larger than 256 bytes. The disk is saturated at 1.7 to 1.8 GByte/s with 2 KB (and larger) chunks. The throughput is even higher in some cases than the maximum throughput specified by the manufacturer showing that not all data has been written to disk when the benchmark was finished. Additionally, the good performance comes at the cost of the page cache using more than 30 GB of the main memory. When limiting the amount of dirty pages, as expected, the performance degrades to around 1 GB/s for large chunks. Increasing the memory pressure does not further degrade the performance in this scenario because the logged data is never read, rendering the read cache useless (it is still larger than 30 GB). Using the RandomAccessFile in synchronous mode has a large penalty on the throughput, which is reduced to around 600 MB/s. When writing to disk with O\_DIRECT, the access is synchronous as well, but the performance is considerable better than the synchronous RandomAccessFile as double buffering is prevented. Actually, up to 256-byte chunks the logging throughput is better than all RandomAccessFile configurations mostly due to the DirectByteBuffer being faster than the HeapByteBuffer (see Section \ref{eval_buffers}). For the targeted chunk sizes of 32 to 256 bytes, DXRAM is able to log more than three million chunks per second, peaking at around 4.64 million 128-byte chunks per second. With around 930 MB/s for 2 to 16 KB chunks the DXRAM's logging performance with O\_DIRECT is equal to copying 8 MB chunks with dd and a single thread (914 MB/s) but is much faster than copying 8 KB chunks with dd (477 MB/s).

\subsubsection{Reorganization}
\label{eval_reorg}
For evaluating the reorganization or more specific the logging performance when the reorganization runs concurrently, we use four different access distributions: sequential, random, zipf and hotNcold (see enumeration below).
The reorganization tests have two phases. First, all chunks are written sequentially to disk (equal to the logging test). Second, twice as many chunks are updated according to the access distribution. For example, when using the sequential distribution, all chunks are written three times in sequential order. With random distribution, on the other hand, all $x$ chunks are written sequentially and then $2*x$ chunks are chosen randomly to be updated. Chunks are written in batches of ten to reduce the overhead for the benchmark itself. The batch size does not affect the logging performance because every chunk needs to be processed solely by the log component (to create a log entry header with unique version, checksum and more).

\begin{enumerate}
	\item Sequential: Updating the chunks in ascending order from first to last in batches of ten. Repeat until number of updates is reached.
	\item Random: Choosing a chunk randomly and update it with the nine following chunks (locality). If the randomly chosen chunk is at the end of a backup zone, the nine preceding chunks are updated. Repeat until the number of updates is reached.
	\item Zipf: Every chunk has an allocated probability to be selected according to the zipf distribution. Select nine succeeding or preceding chunks to complete the batch. Repeat until the number of updates is reached.	
	 
	The zipf distribution follows Zipf's empirical law which is a power law probability distribution studied in physical and social sciences. The zipf distribution allocates the frequency (probability to be chosen) inversely proportional to the rank in the frequency table. The $n$th most likely chosen element has the probability:
	\begin{equation}
	\begin{split}
	& \frac{1}{\sum_{1}^{N}\frac{1}{n^{e}}} * \frac{1}{n^{e}}
	\end{split}
	\end{equation}
	with $e$ being the skew. The benchmark has a freely selectable skew, but we consistently used $1$ (harmonic series) for the evaluation which is close to the distribution in social media networks \cite{zipfsmn}. With the skew $1$, the first element has a probability of nearly 7\% to be chosen, the second 3.5\%, the third 1.7\%.

	Efficiently accessing chunks with zipf distribution requires to generate the distribution before starting phase 2 of the benchmark because calculating the distribution on-the-fly is either too slow or mitigates the distribution. In \cite{zipf}, the authors present a fast method to choose elements according to the zipf distribution without creating the distribution apriori. However, this method allocates the highest probability to the first element, the second highest to the second element and so on. Thus, the values need to be hashed for scrambling the elements. This might destroy the zipf distribution if the hash function does not scatter uniformly (the value range is user-defined). Instead, we create two arrays prior to phase 2 of the benchmark. The first array contains the aggregated frequencies for all chunks, i.e., the value at index $x$ is the probability of choosing the elements 0 to $x - 1$ according to the zipf distribution. The second array is a permutation of all chunk IDs. To choose a chunk, a random value $p$ in $[0.0, 1.0)$ is generated. Afterwards, we search for $p$ or the succeeding value within the first array (binary search). The index $i$ of the searched value is used to index into the second array. Finally, the chunk ID at $i$ is selected to be updated.
	\item HotNcold: Divide all chunks into two partitions: hot and cold. The hot partition contains 10\% of the chunks (the cold 90\%) and 90\% of all updates are chosen from the hot partition (10\% from the cold partition). Select nine succeeding or preceding chunks to complete the batch. Repeat until the number of updates was reached.
	
	We create two arrays prior to phase 2 of the benchmark. The first array has $N * 0.9$ entries and the second $N * 0.1$. We store all cold chunks in the first array and the hot chunks in the second array. Whether a chunk is hot or cold is decided by generating a random value in $[0.0, 1.0)$ for every single chunk. If the value is $< 0.1$ it is considered hot and its chunk ID is stored in the second array. Otherwise, the chunk is cold and stored in the first array.
	
	During phase 2, a chunk is chosen by generating a random value in $[0.0, 1.0)$. If the value is $< 0.9$, we choose a random chunk from the second array (hot), otherwise we choose a chunk from the first array (cold).
\end{enumerate}

\begin{figure}[!t]
	\vspace{-15pt}
	\centering
	\includegraphics[width=3.5in]{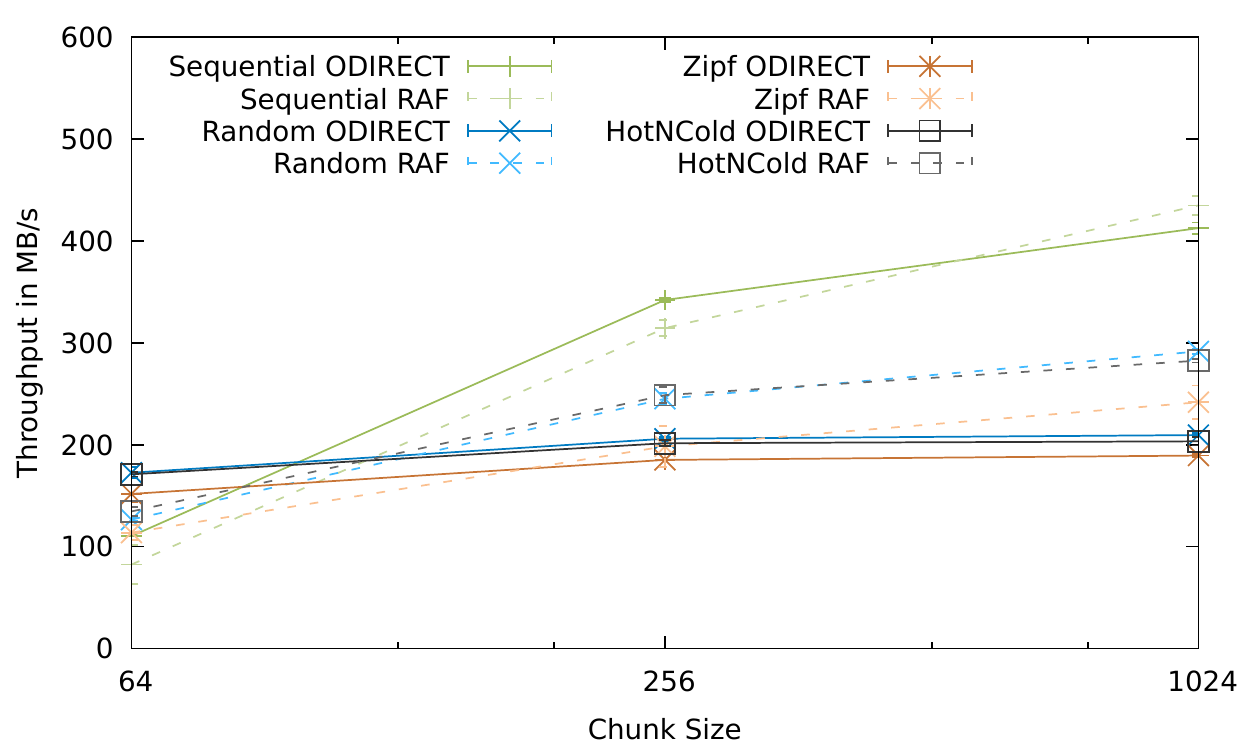}
	\caption{Evaluation of the reorganization. Every chunk is written once, in sequential order. Then, twice as much updates are written according to given distribution}
	\label{reorg_eval_fig}
\end{figure}

Figure \ref{reorg_eval_fig} shows the logging throughput during the reorganization test with 200,000,000 64-byte, 50,000,000 256-byte and 12,500,000 1024-byte chunks stored in 56 backup zones. As chunks are in average updated three times during all runs, the reorganization has to free space for updates to be written. More precisely, during phase 1 the reorganization idles as none of the secondary logs exceed their backup zone size, which indicates that the logs have no invalid data. In phase 2, the reorganization must free at least the amount of chunks written in phase 1 to have enough space in the logs to write all updates. We compare the RandomAccessFile with O\_DIRECT. In contrary to the logging tests, we evaluated the RandomAccessFile with forced writes and memory pressure (between 87.5 and 92.5\% depending on the memory consumption of the distribution), only, because the other configurations are not applicable or too slow for real-world applications. Phase 1 is not included in the throughput measurements.

Again, the direct access is considerably faster for small chunks. For 64-byte chunks, around $2 * 10^{6}$ chunks can be written to disk per second for all distributions. With larger chunks, the RandomAccessFile surpasses O\_DIRECT for all distributions but the sequential distribution. This is because (1) the write accesses for each backup zone are much smaller because they are scattered across all backup zones. Therefore, buffering write accesses in the page cache improves the throughput (but increases the probability of data loss). (2) The page cache stores frequently accessed pages of the disk which can improve the throughput of the reorganization, too. But, this comes at a high price because the page cache puts the system under a high memory pressure which can lead to processes even being killed by the operating system what happened several times during the evaluation. From here on, all tests were executed with O\_DIRECT access as the partly better performance of the RandomAccessFile is outweighed by the problems described before and due to DXRAM being design for very small chunks which are logged/reorganized faster with O\_DIRECT.

For all distributions but the sequential distribution, the performance degradation in comparison to the logging test is caused by the arbitrary access to backup zones which make the aggregation much less efficient. For example, if 200,000,000 64-byte chunks are being accessed randomly, the probability is very high that all 56 backup zones are contained in every flush of the write buffer. During the loading phase (sequential), at most two backup zones are contained producing much larger write accesses. During all runs with random and hotNcold distributions, not one log was filled-up. With the zipf distribution the writer thread was blocked once in a while due to the logs storing the hot spots being full (the two to three most frequently updated chunks). Therefore, the zipf distribution is a little slower than the random and hotNcold distributions. For larger chunks, the three distributions are still restrained by the scattered access. The throughput of the sequential distribution, on the other hand, is dictated by the reorganization throughput. Thus, the throughput is worse for small chunks but improves significantly for larger chunks due to the reorganization being more efficient for larger chunks.

In order to log two million 64-byte chunks per second in phase 2, the reorganization has to free around two million chunks per second as well. With a utilization of 80\% this results in reading 5.33 million and writing 4 million chunks per second. Additionally, version numbers have to be read from disk for the reorganization and written to disk after the reorganization and during the logging. For 64-byte chunks, this are around 3.3 million version numbers per log (without invalid entries).

\textbf{Activation of the Reorganization:} We implemented three mechanisms to activate the reorganization: (1) if a log is larger than a given threshold (e.g., 60\%), it is available for the periodic reorganization, which selects the largest log for reorganization. (2) If the log size exceeds another threshold during writing to it (e.g., 80\%), the writer thread prompts the reorganization by registering a reorganization request for the specific log. The reorganization thread prioritizes requests over the largest log, but finishes reorganizing the currently selected log first. At last, (3) if the writer thread is not able to write to a log because it is full or the fragmentation is too high to write all log entries, the writer registers an urgent reorganization request and awaits its execution. Urgent requests have the highest priority and are processed as soon as possible. 

\begin{figure}[!t]
	\vspace{-15pt}
	\centering
	\includegraphics[width=3.5in]{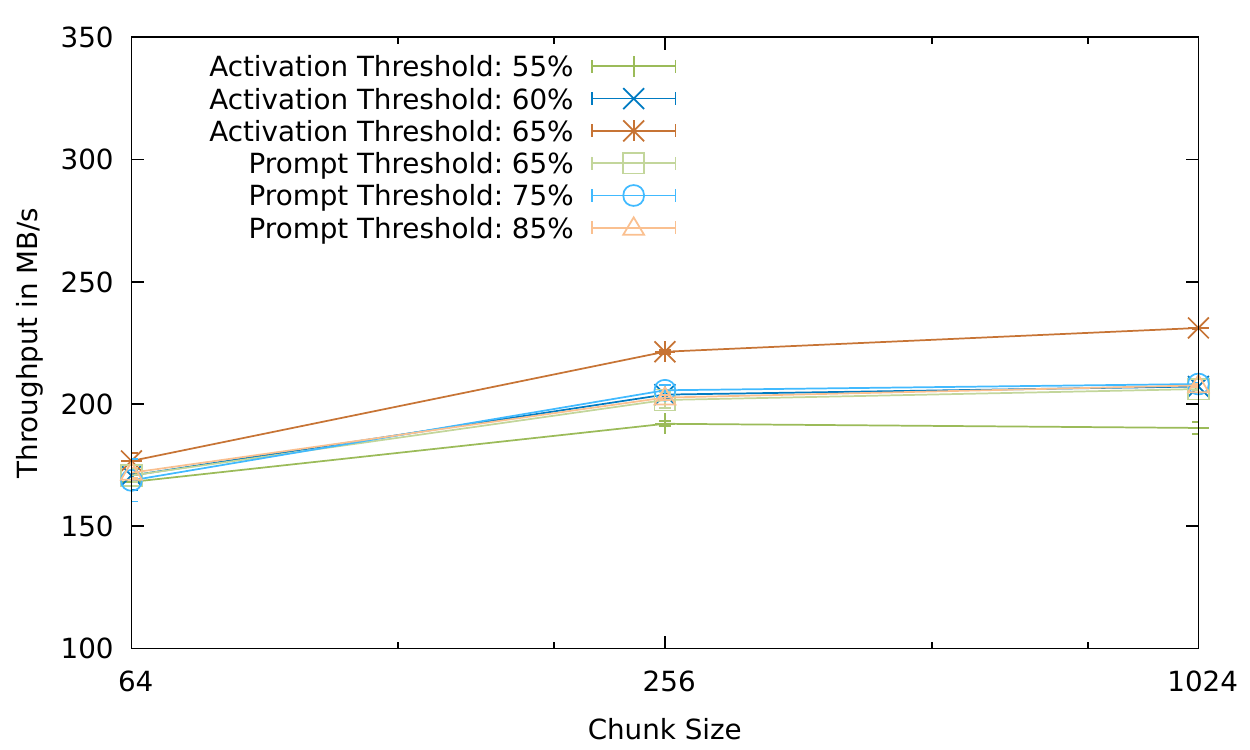}
	\caption{Evaluation of the reorganization thresholds with random distribution.}
	\label{thresholds_eval_fig}
\end{figure}

Figure \ref{thresholds_eval_fig} shows the logging throughput for a random distribution with varying activation (case 1) and prompt (case 2) thresholds. An activation threshold of 65\% and an prompt threshold of 75\% is the best choice in this scenario. With a higher activation threshold (beyond 65\% was not tested), logs are reorganized too late which increases the pressure on the reorganization. If many logs breach the threshold in a short time, the reorganization cannot keep pace. With a lower activation threshold, too much work is done with a low utilization which is not efficient. With 55\% an average of 2.98 MB are freed per reorganized segment, with 65\% 3.95 MB. With a large prompt threshold, the reorganization might miss reorganizing a filling up log. If the prompt threshold is too low, the request queue might grow large and not necessarily the log with most pressure is reorganized but the first reaching the threshold (could still be 65\% whereas another log could be at 95\%, for instance). We uses 60\% activation and 75\% prompt thresholds throughout all other tests.

\begin{figure}[!t]
	\vspace{-15pt}
	\centering
	\includegraphics[width=3.5in]{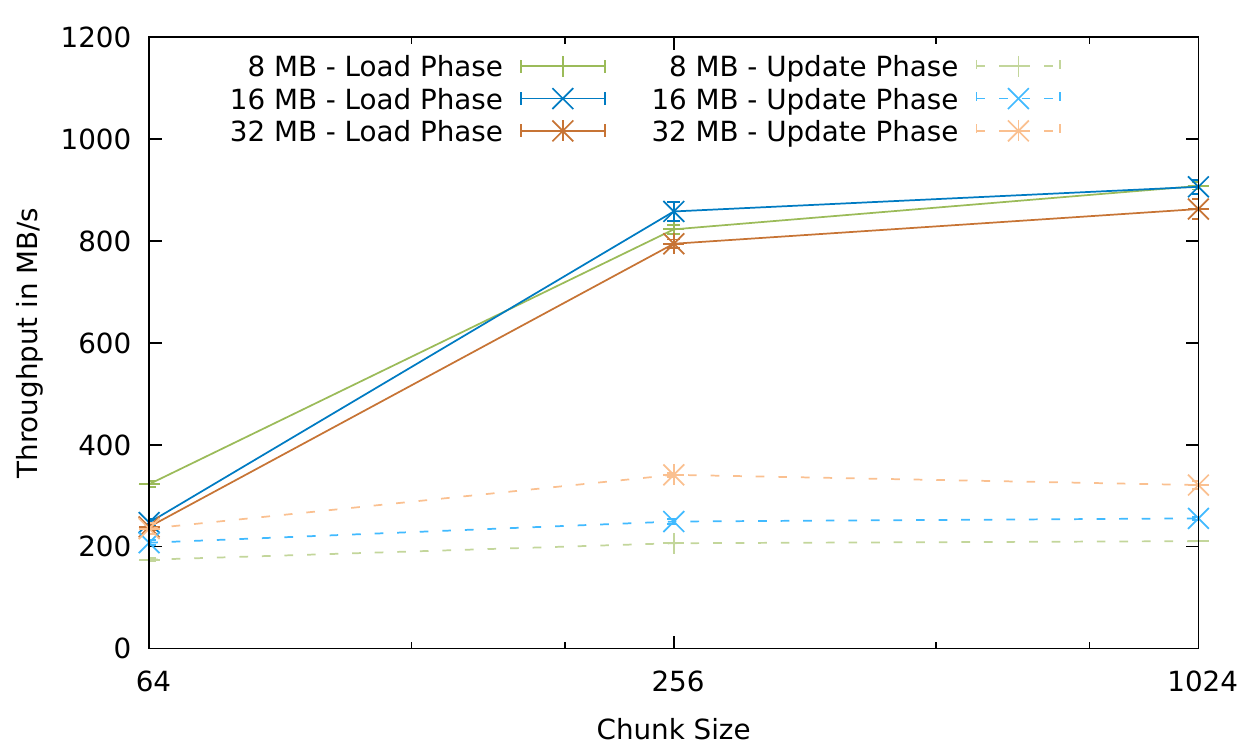}
	\caption{Evaluation of the reorganization with different segment sizes}
	\label{segments_eval_fig}
\end{figure}

\textbf{Segment Size:} We also evaluated the impact of the segment size on the logging and reorganization performance (Figure \ref{segments_eval_fig}). Interestingly, the reorganization benefits from larger segment sizes whereas the logging performance degrades. Larger segments allow the reorganization to process more log entries between I/O accesses which improves the performance. During the loading phase (sequential distribution), write accesses can be aggregated very efficiently because all chunks in the write buffer belong to one or two backup zones, only. However, while the average write access size with 8 MB segments is around 76\% of the segment size (6.06 MB), it is reduced for 32 MB segments to 51\% (17.23 MB; with 16 MB segments: 54\%, 9.06 MB). This results in more often stocking up the larger segments which is slower than writing to the beginning of a segment as the data likely needs to be moved within the buffer. This also affects the reorganization, i.e., segments with higher utilization are beneficial for the reorganization. In all other tests, we use a segment size of 8 MB because it is a good compromise between logging and reorganization performance (especially for small chunks) and it has the lowest memory consumption as pooled buffers are smaller.

\begin{figure}[!t]
	\vspace{-15pt}
	\centering
	\includegraphics[width=3.5in]{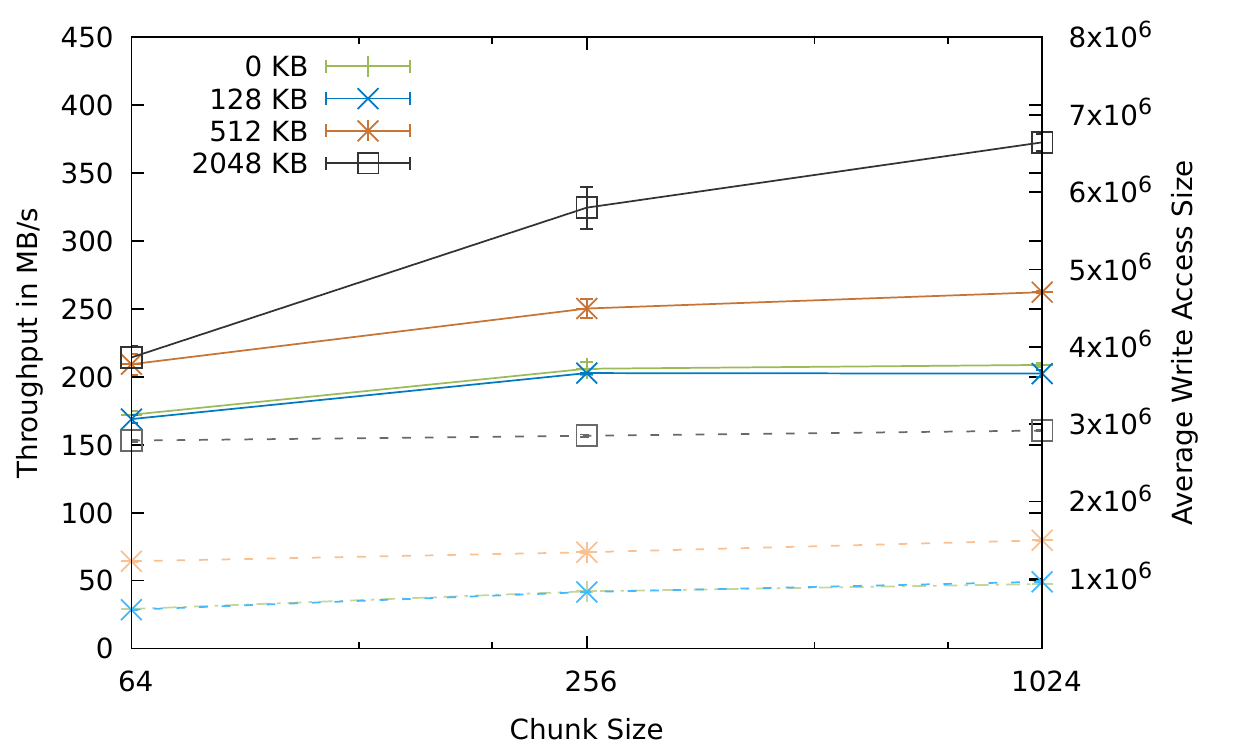}
	\caption{Evaluation of the two-level logging with random distribution and varying secondary log buffer size, 200,000,000 chunks. Solid lines: throughput in MB/s, dashed lines: average write access size}
	\label{two_level_eval_fig}
\end{figure}

\begin{figure}[!t]
	\vspace{-15pt}
	\centering
	\includegraphics[width=3.5in]{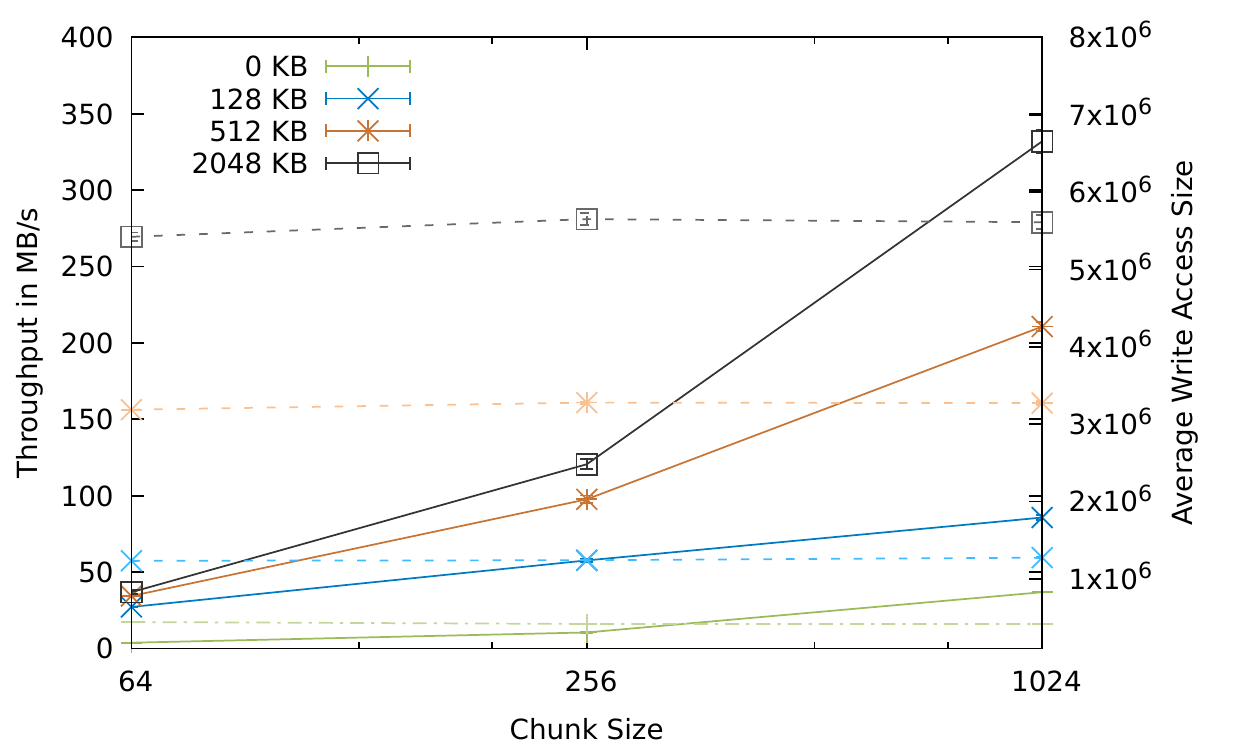}
	\caption{Evaluation of the two-level logging with random distribution and varying secondary log buffer size, 2,000,000,000 chunks. Solid lines: throughput in MB/s, dashed lines: average write access size}
	\label{two_level_large_eval_fig}
\end{figure}

\textbf{Two-Level Logging:} In Figures \ref{two_level_eval_fig} and \ref{two_level_large_eval_fig}, we evaluated the two-level logging by varying the secondary log buffer sizes. The size of the secondary log buffers impacts the logging significantly as it defines the threshold for log entry batches to be written to secondary log or to the primary log and secondary log buffer. For example, if the secondary log buffers have a size of 128 KB, all sorted and aggregated batches from the write buffer smaller than 128 KB are written to primary log and secondary log buffer and all batches equal or larger than 128 KB are directly written to the specific secondary log. With a size of $0$, all log entries are flushed to secondary logs disabling the primary log and secondary log buffers.

Figure \ref{two_level_eval_fig} shows that the two-level logging with secondary log buffer sizes larger than 512 KB increases the throughput by 20 to 25\% for the random distribution and 200,000,000 64-byte chunks and up to 117\% for 1024-byte chunks. This is due to the write accesses being much larger (dashed lines in Figure \ref{two_level_eval_fig}). However, using very large secondary log buffers increases the wear on the disk as many log entries are written twice (first to primary log, later to secondary log). Furthermore, the data processing is more time consuming than writing to disk in this scenario. This would not be the case for slower disks making smaller secondary log buffers more attractive. In this workload, using 128 KB secondary log buffers is as fast as disabling the two-level logging because the average batch size in the write buffer is considerably larger than 128 KB. 

To decrease the log entry batch sizes, we repeated the test from above with 2,000,000,000 64-byte chunks (500,000,000 256-byte and 125,000,000 1024-byte chunks). Figure \ref{two_level_large_eval_fig} shows that the performance advantage of utilizing the two-level logging increases with more chunks and thus smaller batch sizes when using the random distribution, as expected. With 128 KB secondary log buffers, the two-level logging improves the performance for 64-byte chunks by more than seven times in comparison to a normal logging scheme.

The random distribution is the worst case scenario regarding the decrease of batch sizes with increasing number of chunks because it scatters the accesses uniformly across all logs making aggregation less efficient with many chunks. The sequential distribution is unaffected by the number of chunks, the zipf and hotNcold distributions are less affected than the random distribution.

\subsection{Timestamps}
\begin{figure}[!t]
	\vspace{-15pt}
	\centering
	\includegraphics[width=3.5in]{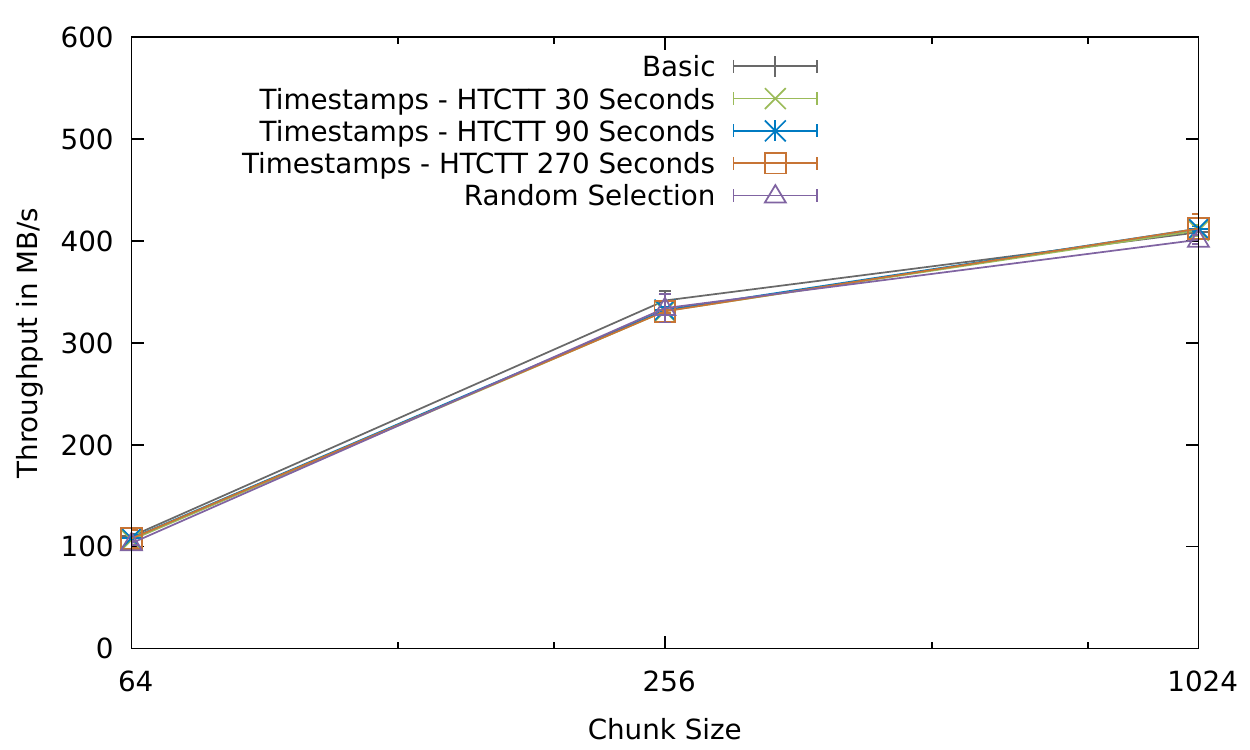}
	\caption{Evaluation of the reorganization with timestamps. Sequential access distribution}
	\label{timestamps_seq_eval_fig}
\end{figure}

\begin{figure}[!t]
	\vspace{-15pt}
	\centering
	\includegraphics[width=3.5in]{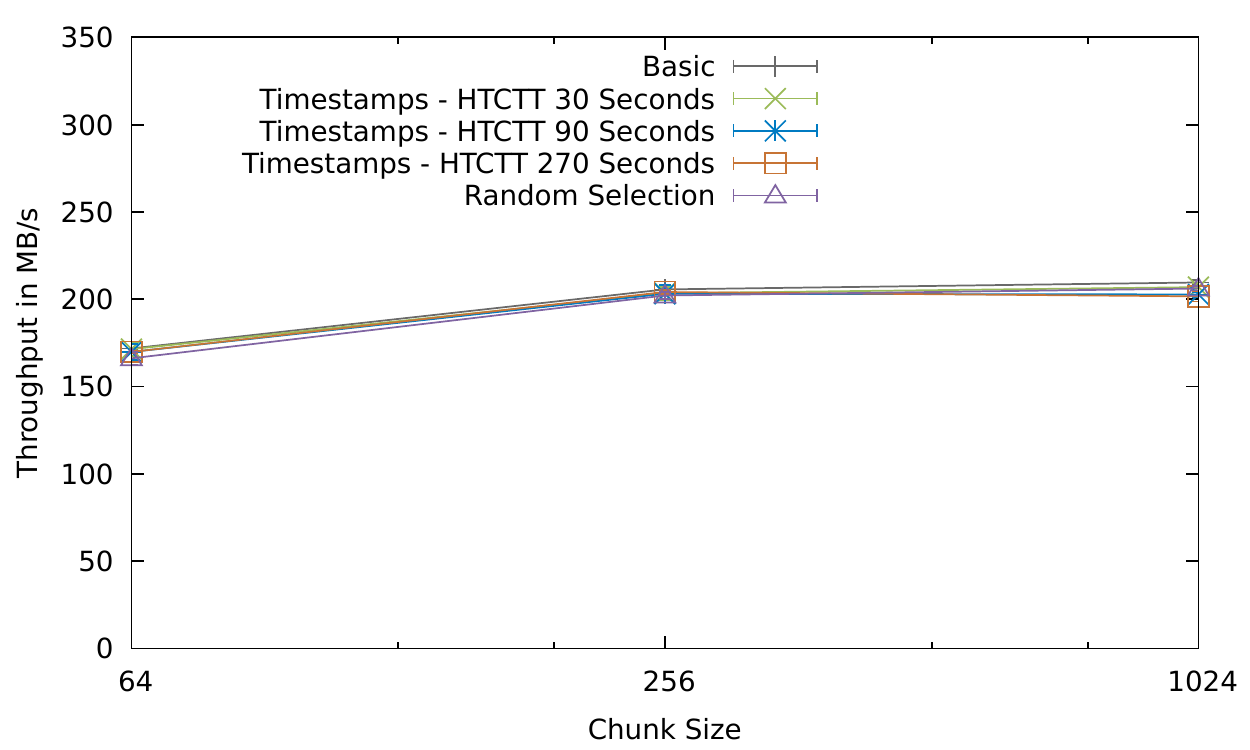}
	\caption{Evaluation of the reorganization with timestamps. Random access distribution}
	\label{timestamps_rand_eval_fig}
\end{figure}

\begin{figure}[!t]
	\vspace{-15pt}
	\centering
	\includegraphics[width=3.5in]{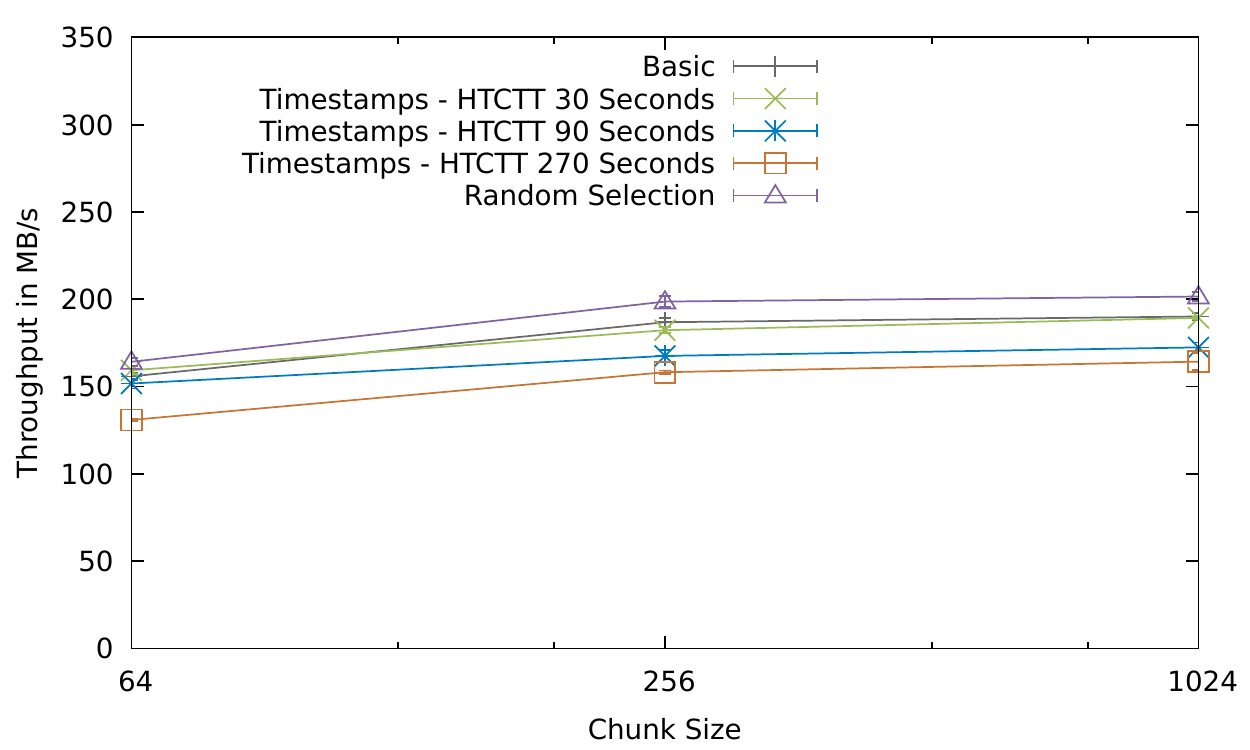}
	\caption{Evaluation of the reorganization with timestamps. Zipf access distribution}
	\label{timestamps_zipf_eval_fig}
\end{figure}

\begin{figure}[!t]
	\vspace{-15pt}
	\centering
	\includegraphics[width=3.5in]{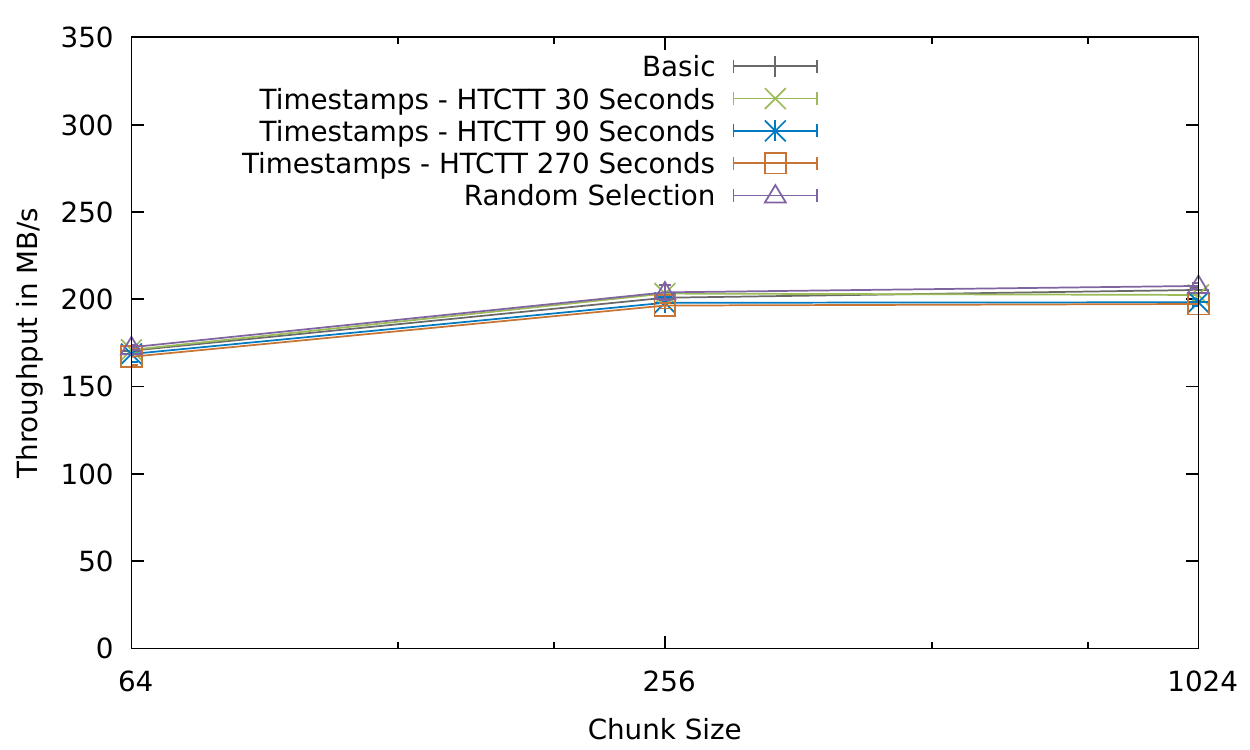}
	\caption{Evaluation of the reorganization with timestamps. HotNCold access distribution}
	\label{timestamps_hnc_eval_fig}
\end{figure}

Figures \ref{timestamps_seq_eval_fig} to \ref{timestamps_hnc_eval_fig} show the logging throughput with ongoing reorganization. In contrary to Section \ref{eval_reorg}, we used three different segment selection strategies: basic ($time$ $since$ $last$ $reorganization$ $or$ $creation$ $*$ $utlization$), with timestamps to determine the average age of a segment ($age$ $*$ $utilization$) and random selection. We also varied the hot-to-cold transformation threshold (HTCTT) to study its impact on the performance.

Surprisingly, Figures \ref{timestamps_seq_eval_fig} to \ref{timestamps_hnc_eval_fig} show that the segment selection has a negligible impact on the overall performance. For the sequential, random and hotNcold distributions all five selection strategy are equal regarding the logging throughput with ongoing reorganization. Only, for the zipf distribution, the throughput differs. However, the least elaborated strategy has the highest throughput in this scenario. 

The timestamp selection has to be considerably better than the other strategies to outweigh the additional four bytes in the log entry headers. This is not the case here. For the random and hotNcold distribution the reorganization is not under pressure because the logging is restrained by the scattered access. Therefore, the segment selection cannot make a difference in this scenario. For the sequential distribution, the logging throughput is not restrained, but logs fill-up one after another, quickly triggering urgent requests for the current log. An urgent request initiates a reorganization of all segments in ascending order rendering the segment selection strategy irrelevant. For the zipf distribution, selecting older segments can be misleading because new segments of logs with a hotspot contain many already outdated versions of the very frequently updated hotspot. Hence, selecting a new segment for the reorganization can be better in this scenario. A low HTCTT has a positive affect on the segment selection as older segments appear much younger (older objects are left out for the age determination), in some cases even younger than a new segment.

\subsection{Logging Remote Chunks}
\label{eval_remote}
In this section, we evaluate the logging performance with chunks transferred over an InfiniBand network. We used the O\_DIRECT access, 8 MB segments, the two-level logging with 128 KB threshold and no timestamps. The checksums are used like in all other tests. The benchmark creates all chunks (up to 400,000,000), first. Then, the chunks are updated sequentially in batches of ten which are sent directly over the network to the backup server.

\begin{figure}[!t]
	\vspace{-15pt}
	\centering
	\includegraphics[width=3.5in]{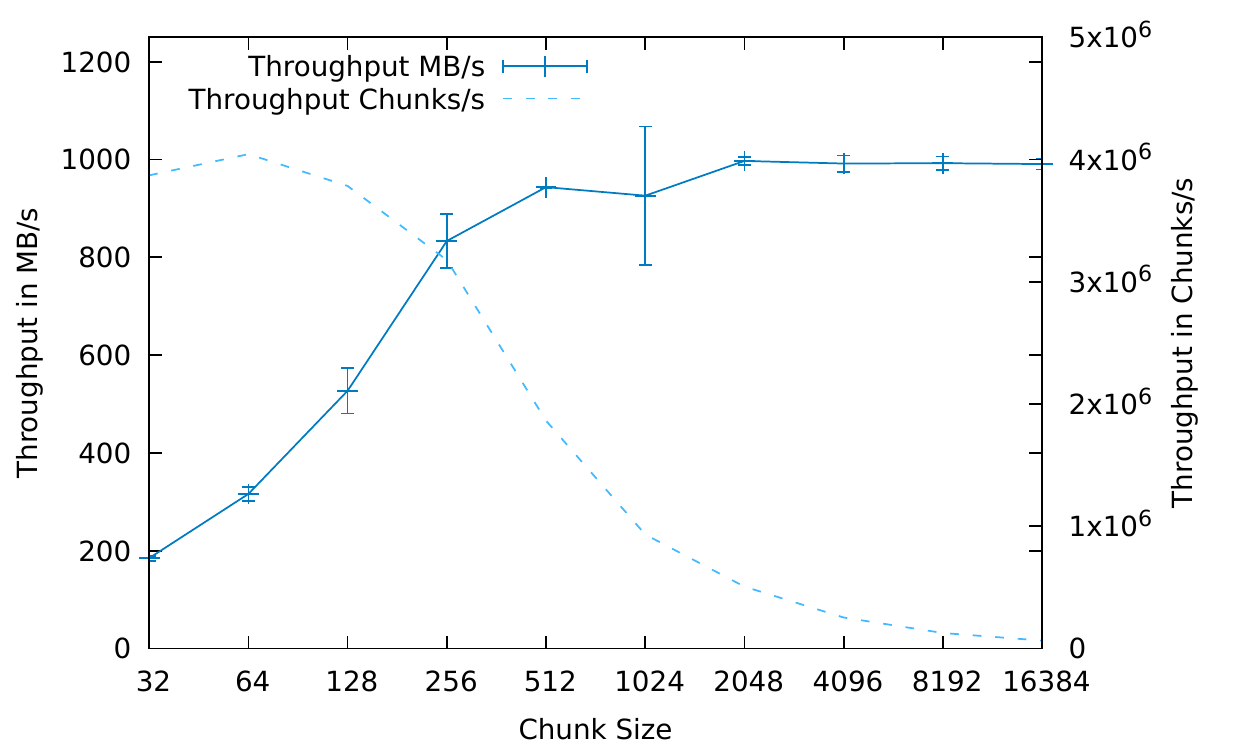}
	\caption{Logging Throughput over InfiniBand Network}
	\label{net_eval_fig}
\end{figure}

Figure \ref{net_eval_fig} shows that no performance is lost when chunks are sent over the network instead of creating and logging them locally. DXRAM is able to update, sent, receive and log more than 4,000,000 64-byte chunks per second. The SSD is saturated with up to 512-byte chunks with a throughput of nearly 1 GB/s.

\section{Conclusions}
\label{conclusions}
In this report, we presented DXRAM's logging architecture in detail with focus on the data flow and the disk access methods, extending the papers \cite{dxram4} and \cite{dxram5}. Furthermore, we discussed the usage of timestamps to accurately calculate a segments age in order to improve the segment selection for the reorganization and we introduced copysets to DXRAM.

The evaluation shows the good performance of the logging and reorganization and demonstrates that DXRAM utilizes high throughput hardware like InfiniBand networks and nvme PCIe SSDs efficiently. DXRAM is able to log more than 4,000,000 64-byte chunks per second received over an InfiniBand network. Larger chunks, e.g., 512-byte chunks, can be logged at nearly 1 GB/s, saturating the PCI-e SSD. The reorganization is able to keep the utilization most times under 80\% for realistic distributions (random, zipf and hotNcold) while maintaining a high logging throughput.

\bibliographystyle{IEEEtran}
\bibliography{references}

\end{document}